\documentclass[aps,twocolumn,superscriptaddress]{revtex4-1}
\usepackage{graphicx}
\usepackage{longtable,array}
\usepackage{amsmath}
\usepackage{color}
\usepackage{float}
\usepackage{hyperref}
\usepackage{wasysym}
\usepackage{url}
\usepackage{dcolumn}
\usepackage{bm}

\usepackage[T1]{fontenc} 

\draft
\begin{document}

\title{Leveraging resonant frequencies of an optical cavity for spectroscopic measurement of gas temperature and concentration}

\author{Daniel Lisak}
\email{dlisak@umk.pl}
\affiliation{Institute of Physics, Faculty of Physics, Astronomy and Informatics, Nicolaus Copernicus University in Toru\'n, Grudziadzka 5, 87-100 Torun, Poland}
\author{Vittorio D'Agostino}
\affiliation{Institute of Physics, Faculty of Physics, Astronomy and Informatics, Nicolaus Copernicus University in Toru\'n, Grudziadzka 5, 87-100 Torun, Poland}
\affiliation{Department of Mathematics and Physics, Universit\`a degli Studi della Campania ”Luigi Vanvitelli”, Caserta, 81100, Italy}
\author{Szymon W\'{o}jtewicz}
\affiliation{Institute of Physics, Faculty of Physics, Astronomy and Informatics, Nicolaus Copernicus University in Toru\'n, Grudziadzka 5, 87-100 Torun, Poland}
\author{Agata Cygan}
\affiliation{Institute of Physics, Faculty of Physics, Astronomy and Informatics, Nicolaus Copernicus University in Toru\'n, Grudziadzka 5, 87-100 Torun, Poland}
\author{Marcin Gibas}
\affiliation{Central Office of Measures, Physical Chemistry and Environment Department, Świętokrzyskie Laboratory Campus of the Central Office of Measures, Poland, Poland}
\author{Piotr Wcis{\l}o}
\affiliation{Institute of Physics, Faculty of Physics, Astronomy and Informatics, Nicolaus Copernicus University in Toru\'n, Grudziadzka 5, 87-100 Torun, Poland}
\author{Roman Ciury{\l}o}
\affiliation{Institute of Physics, Faculty of Physics, Astronomy and Informatics, Nicolaus Copernicus University in Toru\'n, Grudziadzka 5, 87-100 Torun, Poland}
\author{Katarzyna Bielska}
\affiliation{Institute of Physics, Faculty of Physics, Astronomy and Informatics, Nicolaus Copernicus University in Toru\'n, Grudziadzka 5, 87-100 Torun, Poland}

\date{\today}

\begin{abstract}
We introduce a spectroscopic approach to primary gas thermometry, harnessing precise optical cavity resonance frequencies and ab initio molecular line intensity calculations. By utilizing CO (3-0) vibrational band lines and cavity mode dispersion spectroscopy, we achieve an uncertainty of 82 ppm (24 mK at 296 K) in line-intensity-ratio thermometry (LRT)—over an order of magnitude lower than any previously reported spectroscopic thermometry at gas pressures above 1.2 kPa. This method extends high-precision spectroscopic thermometry across a pressure range an order of magnitude larger than prior techniques, enabling a fully optical, non-contact, and molecule-selective primary amount-of-substance measurement. We further demonstrate sub-permille uncertainty in gas concentration measurements across pressures from 50 Pa to 20 kPa, significantly enhancing the precision and versatility of spectroscopic gas metrology.

\end{abstract}

\maketitle

\section{Introduction}
\label{sec1}

The redefinition of the International System of Units (SI) \cite{Stock2019} separated the unit definitions, based on a set of fixed physical constants, from their realizations, giving us flexibility in choosing suitable physical equations linking the defining constants to the quantity we wish to measure. In the case of temperature, the accurate methods used before to determine the Boltzmann constant $k$ from the thermal energy, $kT$, are now used for the primary temperature, $T$, measurement \cite{Fellmuth2016}.
Among optical kelvin realizations in the gas phase, the Doppler broadening thermometry (DBT) \cite{Daussy2007, Gianfrani2016} and the rovibrational line intensity ratio thermometry (LRT) (also called rotational-state distribution thermometry) \cite{Arroyo1993, Berman1993} are promising methods in terms of accuracy \cite{Shimizu2018, SantamariaAmato2019, Gotti2020}. 
The DBT relies on the measurement of the Doppler broadening of molecular lines and, therefore, achieves the highest accuracies at very low gas pressures, at which the spectral line shape is only weakly affected by collisional effects. With increasing pressure, the line shape becomes complicated \cite{Ciurylo2001JQSRT, Ciurylo2002BB, Ngo2013}, and approximate line-shape models fitted to the spectra lead to non-negligible systematic errors in the retrieved Doppler widths \cite{Cygan2010, Triki2012, Moretti2013, Wcislo2013}. The Doppler half-width $\Gamma_{\rm D}$ is proportional to $\sqrt{T}$, therefore, its relative sensitivity to temperature change, $\eta_{\rm DBT}=(d\Gamma_{\rm D}/dT)/(\Gamma_{\rm D}/T)$, equals $\frac{1}{2}$. In contrast, the integrated spectral line area used in LRT can be accurately retrieved even in the case of significant contribution of pressure-induced line-shape effects. 
Moreover, an advantage of the LRT over DBT is that the temperature sensitivity of the line intensity ratio, $R_S = S_b/S_a$, can be significantly enhanced by choosing a line pair ($a$ and $b$) with high lower-state rotational energy difference, $\Delta E'' = E''_a - E''_b$. The $R_S$ depends on $T$ through the dominating Boltzmann factor, $\exp[-\Delta E''/(kT)]$ \cite{Rothman1996}. Therefore, the relative sensitivity, $\eta$, defined as a ratio of fractional changes in line intensity pair ratio and temperature $T$, is 
\begin{equation}
    \label{Eqeta}
    \eta= \frac{dR_S/dT}{R_S/T} \approx -\frac{\Delta E''}{kT} \; .
\end{equation}
This sensitivity $\eta$ can be more than an order of magnitude higher than that for DBT. 
The formula for the gas temperature determined from the intensity ratio of two rovibrational lines \cite{Arroyo1993} can be written as
\begin{equation}
    \label{EqT2L}
    T=\frac{\Delta E''}{k}\; \frac{1}{\ln[R_S/\xi(T)]} \;.
\end{equation}
In the case of overtone transitions used for the determination of gas temperature at a few hundred K or less,
$\xi(T)$ (see Appendix \ref{appxi}) is approximated with high accuracy \cite{SantamariaAmato2018} by a constant value specific for the used molecular system
$\xi \approx R_S(T_r)\exp[-\Delta E''/(k T_r)]$. The reference line intensity ratio, $R_S(T_r) = S_b(T_r)/S_a(T_r)$, at arbitrarily chosen reference temperature $T_r$, can be calculated {\it ab initio} for simple molecules \cite{Pachucki2008, Polyansky2015, Bielska2022}. One of the accuracy limitations of the LRT is the uncertainty of {\it ab initio} calculations of $R_S(T_r)$, but high sensitivity factor $\eta$ increases the accuracy of determined $T$ for given accuracies on not only the measured $R_S$ but also the reference $R_S(T_r)$.

While the line intensity ratio can be used for spectroscopic thermometry, the absolute line intensity is commonly used for the spectroscopic amount of substance measurement, based on the relation between the gas concentration, $n$, integrated line area (per unit absorption path length), $A$, and line intensity, $S(T)$, in the linear-absorption regime
\begin{equation}
    \label{Eqn}
    n=A/S(T)\;. 
\end{equation}
We can use Eq. (\ref{Eqn}) because non-impact \cite{Boulet2004, Reed2023} and line mixing effects \cite{Ciurylo2001LM, Pine2019} are negligible (see Appendix \ref{appNILM}) compared to the accuracy reached in our study.
For a multi-component gas mixture, Eq. (\ref{Eqn}) enables selectivity to measure $n$ of a given component using its spectral line. Importantly, spectroscopic temperature measurement is necessary for fully spectroscopic gas concentration measurement because of the temperature dependence of $S$ and, in consequence, enables optical pressure determination using the gas equation of state. 
Moreover, the self-calibrating spectroscopic methods, such as cavity ring-down \cite{Romanini1997}, cavity buildup \cite{Cygan2021}, cavity mode width \cite{Cygan2013},  cavity mode dispersion \cite{Cygan2015}, and heterodyne cavity ring-down \cite{Cygan2025}, eliminate the need for the absorption path length calibration.  

The most accurate spectroscopic primary thermometry results were done at gas pressures below 1 kPa as reviewed in Ref. \cite{Gotti2021}. The relative combined standard uncertainties of $u_r(T)$ below 100 parts-per-million (ppm) were achieved for DBT using a single molecular \cite{Moretti2013, Hashemi2014, Gotti2018} or atomic \cite{Truong2015} line and using a multi-line DBT \cite{Gatti2013, Castrillo2019}. The LRT achieved so far $u_r(T)$ of 2000~ppm using dual-comb spectroscopy of C$_2$H$_2$ \cite{Shimizu2018} and 530~ppm with continuous-wave laser absorption spectroscopy of CO$_2$ \cite{Gotti2020}. However, theoretical considerations \cite{SantamariaAmato2018, Gotti2020} suggested that the uncertainty of tens of ppm or better could be achievable for CO and CO$_2$. In particular, CO is very attractive for spectroscopic thermometry due to its simple molecular structure, well-isolated spectral lines, and noticeably higher concentration stability in spectroscopic cells than CO$_2$. These features are highly advantageous for accurate line intensity measurements and calculations. Indeed, the most accurate line intensities have been so far obtained for the (3-0) band of $^{12}$C$^{16}$O molecule, for which sub-permille combined standard uncertainties and comparable agreement between different experimental and {\it ab initio} intensities near $T= 296$ K were reported for many lines \cite{Bielska2022, CCQMP229, Huang2024}. 

\begin{figure}[t]
\centering
\includegraphics[width=0.48\textwidth]{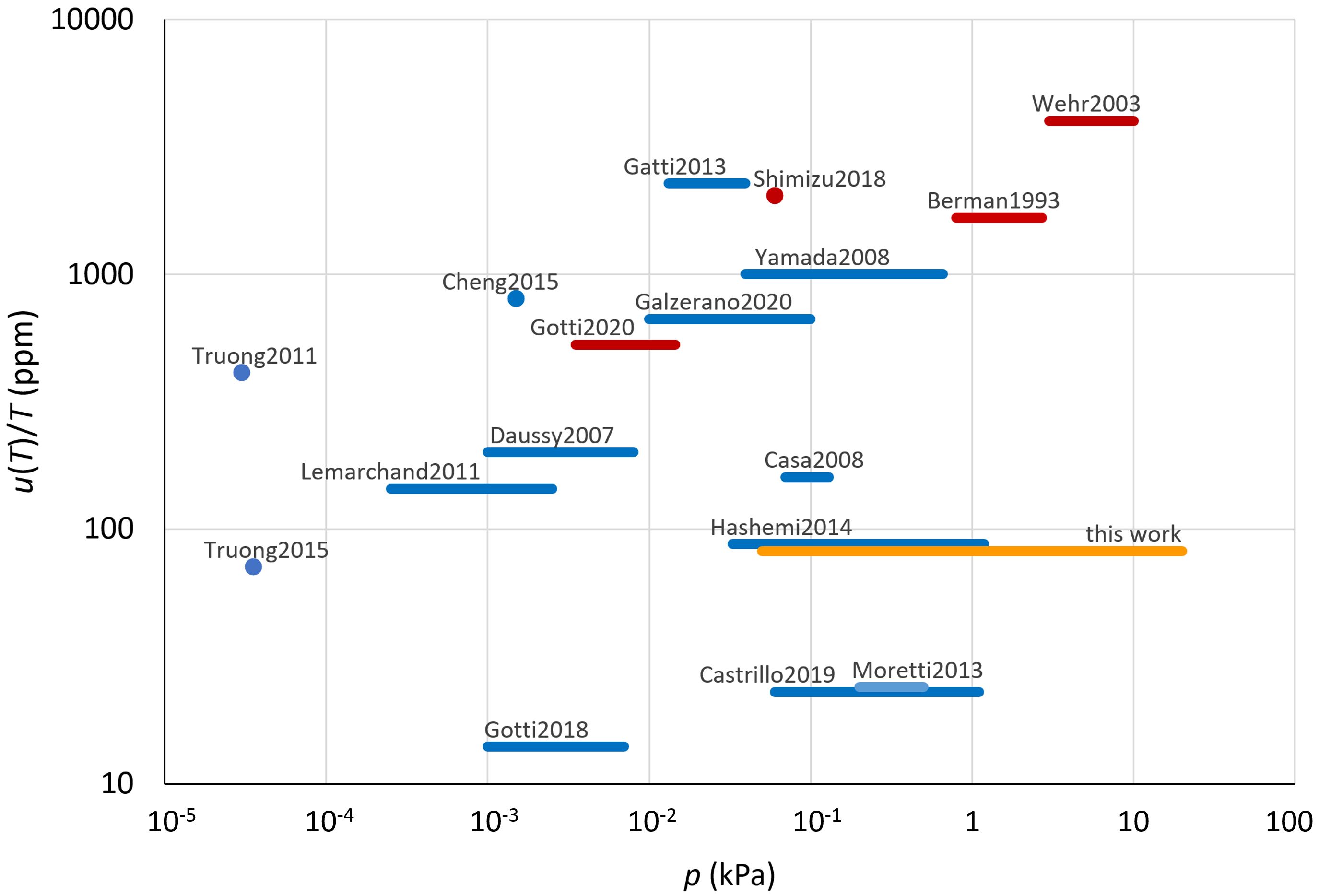}
\caption{The relative uncertainty of spectroscopic temperature plotted against demonstrated gas pressure range in literature data for DBT (blue) \cite{Daussy2007, Casa2008, Yamada2008, Lemarchand2011, Truong2011, Gatti2013, Moretti2013, Hashemi2014, Truong2015, Cheng2015, Gotti2018, Castrillo2019, Galzerano2020}, LRT (red) \cite{Berman1993, Wehr2003, Shimizu2018, Gotti2020}, and this work's frequency-based LRT (orange).} \label{Fig_LiterRev}
\end{figure}

In this Letter, we report a spectroscopic primary gas temperature and concentration measurement method that exploits an optical cavity's resonant frequencies instead of conventional for LRT and DBT light intensity absorption detection. 
This frequency-based spectroscopic approach assures immunity to nonlinearities in the light-intensity detection system leading to a large dynamic range and sub-permille level of potential line-shape distortions \cite{Cygan2019, Bielska2022}.
Using the integrated line area ratios of the CO (3-0) band lines measured with 
a frequency-based cavity mode dispersion spectroscopy (CMDS) \cite{Cygan2015, Cygan2019} at pressures that maximize the accuracy for each line area and {\it ab initio} line intensities reported in Ref. \cite{Bielska2022}, we demonstrate temperature uncertainty of 24 mK at 296 K ($u_r(T)=82$ ppm).
Comparison with the previous DBT and LRT realizations, shown in Fig. \ref{Fig_LiterRev}, reveals that no competitive spectroscopic thermometry was demonstrated for the gas pressure above 1.2 kPa, where we obtained twenty-fold lower uncertainty than previously reported results. Our approach to LRT results in 6.5-fold higher temperature accuracy than any previous LRT and over an order of magnitude larger maximum pressure and pressure span than any DBT. These features make it suitable not only for primary contact-free thermometry applications but also for fully optical amount of substance and pressure measurements. 
Using our frequency-based LRT and integrated line areas of four selected lines, we demonstrate absolute, fully frequency-based spectroscopic gas concentration measurement with permille-level relative standard uncertainties at a pressure range from 50 Pa to 20 kPa. 
This result opens the path to the fully optical, non-contact primary amount of substance metrology, based on measurements of resonant frequencies of the optical cavity and quantum properties of molecules - the spectral line intensities.

\section{Experiment}
\label{sec2}

For the high accuracy of the relative and absolute integrated line area measurements, we use the cavity mode dispersion spectroscopy (CMDS), which has a self-calibrating optical path length, and both the horizontal and vertical axes of the dispersive line shapes are obtained from measurements of the cavity resonant frequencies. 
The CMDS spectrometer is described in Appendix \ref{app1}.
The measured cavity mode frequencies provide the dispersion spectrum of the measured molecular line. The dispersive mode frequency shift, $\Delta\nu_d$, due to the absorption line having intensity, $S$, is \cite{Cygan2015, Cygan2025}
\begin{equation}
    \label{EqDispShift}
    \Delta\nu_d(\nu-\nu_0) = \frac{1}{4\pi n_0}\; n\; S(T)\; \Phi_{\rm I}(\nu-\nu_0)
\end{equation}
where $\nu_0$ is the molecular line center frequency, $n_0$ is the broadband sample refractive index, $T$ is the sample temperature, $\Phi_{\rm I}(\nu-\nu_0)$ is the imaginary part of the complex spectral line-shape function.

The entire cavity is placed in a thermal insulation box with active temperature stabilization at 296 K. 
The combined standard uncertainty of the cavity temperature, $u(T)=30$ mK (see Appendix \ref{app1} for details). The gas pressure,$p$, was measured with capacitance manometers calibrated to SI with a relative combined standard uncertainty $u(p)/p = 0.1$\% to 0.008\% for $p$ of 1.3 to 20~kPa and $u(p)/p= 0.05$\% for $p$ of 20 Pa to 1.3~kPa.

To retrieve integrated line areas, we fitted the CMDS spectra of CO (3-0)-band lines with the Hartmann-Tran lineshape model (HTP) \cite{Ngo2013, Tennyson2014}, which shows good agreement with the experimental spectra (see Appendix \ref{app1}).

\section{Results and discussion}
\label{sec1}

\subsection{Sensitivity of line intensity ratio to temperature}

\begin{figure}[t]
\centering
\includegraphics[width=0.48\textwidth]{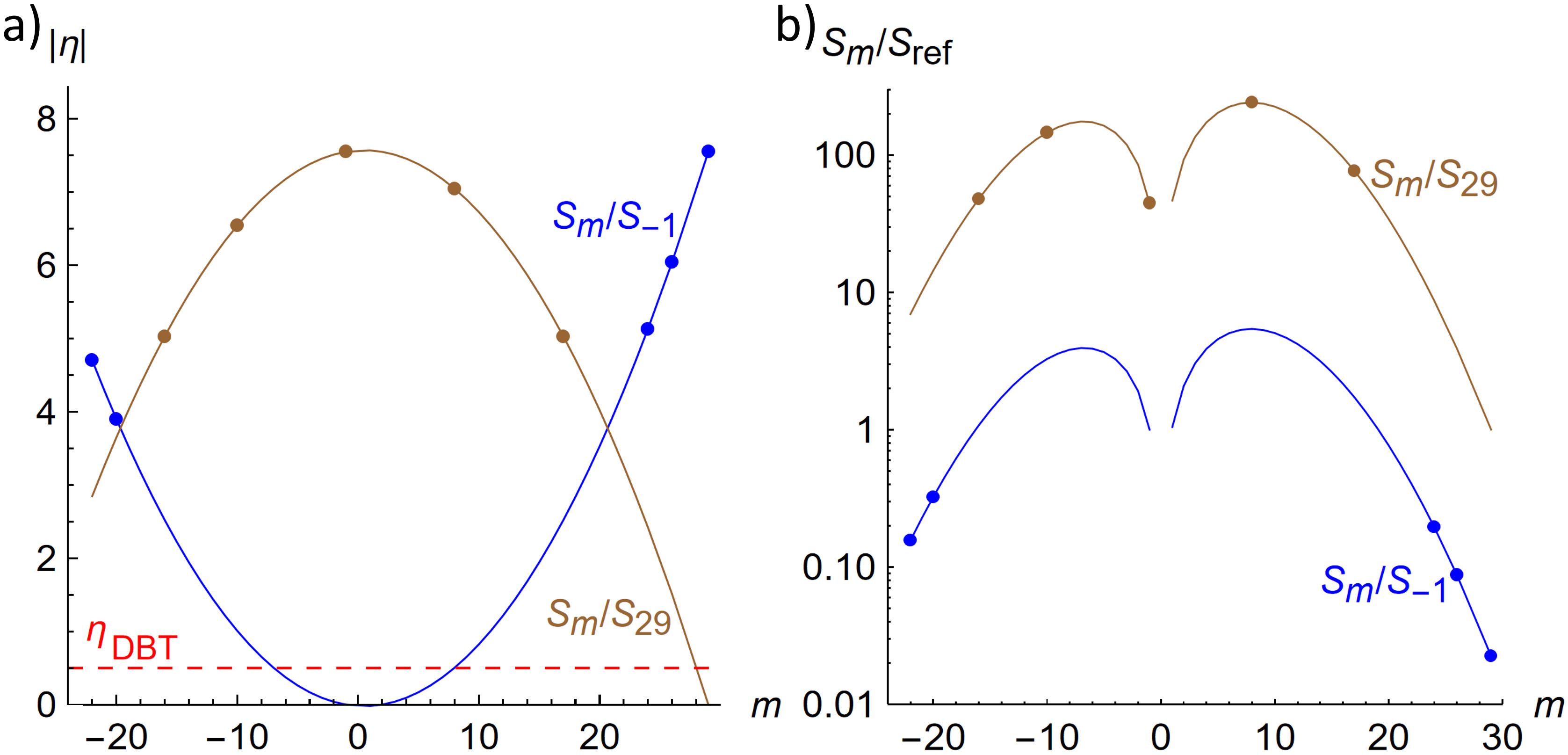}
\caption{a) Sensitivity of CO (3-0) band line intensity ratios $S_{m}/S_{m_2}$ to temperature change near 296 K for $m_2$ set to -1 (blue plot) and 29 (brown plot). For comparison, the sensitivity of DBT, $\eta_{DBT}$, is shown with a red dashed line. b) Ratios $S_{m}/S_{m_2}$ for $m_2$ set to -1 (blue plot) and 29 (brown plot) at 296 K. Points indicate lines measured in this study.} \label{fig_Tsens}
\end{figure}

The sensitivity factor $\eta$, defined by Eq. (\ref{Eqeta}), is shown in Fig. \ref{fig_Tsens}(a) for the CO (3-0) band lines at 296 K versus the line index $m$ for two cases corresponding to line pairs $(m,-1)$ and $(m,29)$. The $m$ index equals $-J''$ and $J''+1$ for the P and R branch lines, respectively, where $J''$ is the lower-state rotational quantum number.
To minimize the uncertainty in our LRT, we use ratios of given line intensity to the line (either -1 or 29), which leads to higher sensitivity $\eta$. From Fig. \ref{fig_Tsens}(b) of the corresponding line intensity ratios, it is clear that enhanced sensitivity, $\eta$, is achieved for lines having significantly different intensities, which is disadvantageous for simultaneous high-accuracy line area measurement. Therefore, we measured the intensities of each selected line at the same $T$ and several CO pressures optimized for the highest SNR of spectra. Note that since LRT operates on line intensity ratios, we only need linear, but not necessarily absolute calibrated CO pressure measurement. This linearity was verified spectroscopically with a standard deviation of 0.023\% for pressures between 50 Pa and 20 kPa (see Appendix \ref{app2}). The selected line pairs provide the temperature sensitivity enhancement factor, $\eta$, up to 7.6 for the intensity ratio of 45. However, $\eta=2.5$ can be achieved for lines having equal intensities.

\subsection{Comparison of spectroscopic and reference gas temperature}

\begin{figure}[t]
\centering
\includegraphics[width=0.48\textwidth]{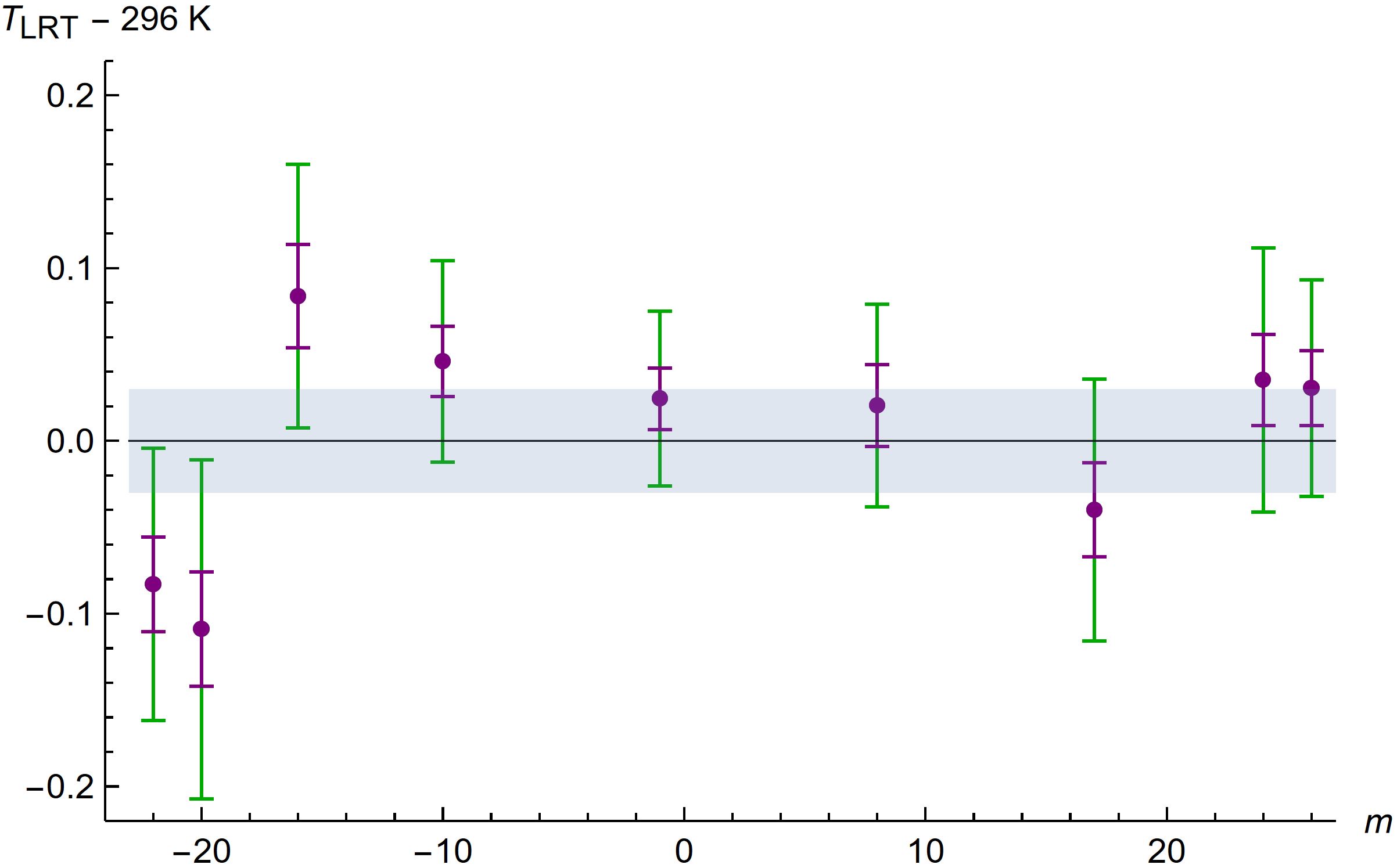}
\caption{Differences between temperatures from LRT, $T_{LRT}$, and the reference gas temperature (296.00(3) K) obtained for line pairs shown in Fig. \ref{fig_Tsens}. Green and purple error bars correspond to combined and statistical standard uncertainties, respectively. The gray area corresponds to the standard uncertainty of $T$ measured with a calibrated reference sensor.} \label{fig_Tspec}
\end{figure}

In Fig. \ref{fig_Tspec}, we show the difference between our measured spectroscopic temperature and the measured temperature of a gas cell with thermistors calibrated to SI. The green uncertainty bars represent combined standard uncertainties, and their inverse squares are used as weights for the mean calculation. The purple uncertainty bars show uncertainties related only to spectroscopic measurements. The weighted mean spectroscopic temperature and its standard deviation are (296.000 $\pm$ 0.024) K, corresponding to a relative standard uncertainty of 82 ppm, while the standard deviation for a single line pair is 64 mK. We note that the P-branch lines lead to a higher scatter of $T$ than the R-branch ones, which is consistent with a reported comparison of the absolute line intensities between experiments and theory \cite{Bielska2022, CCQMP229}.

\subsection{Comparison of spectroscopic and reference gas concentration}

\begin{figure}[t]
\centering
\includegraphics[width=0.48\textwidth]{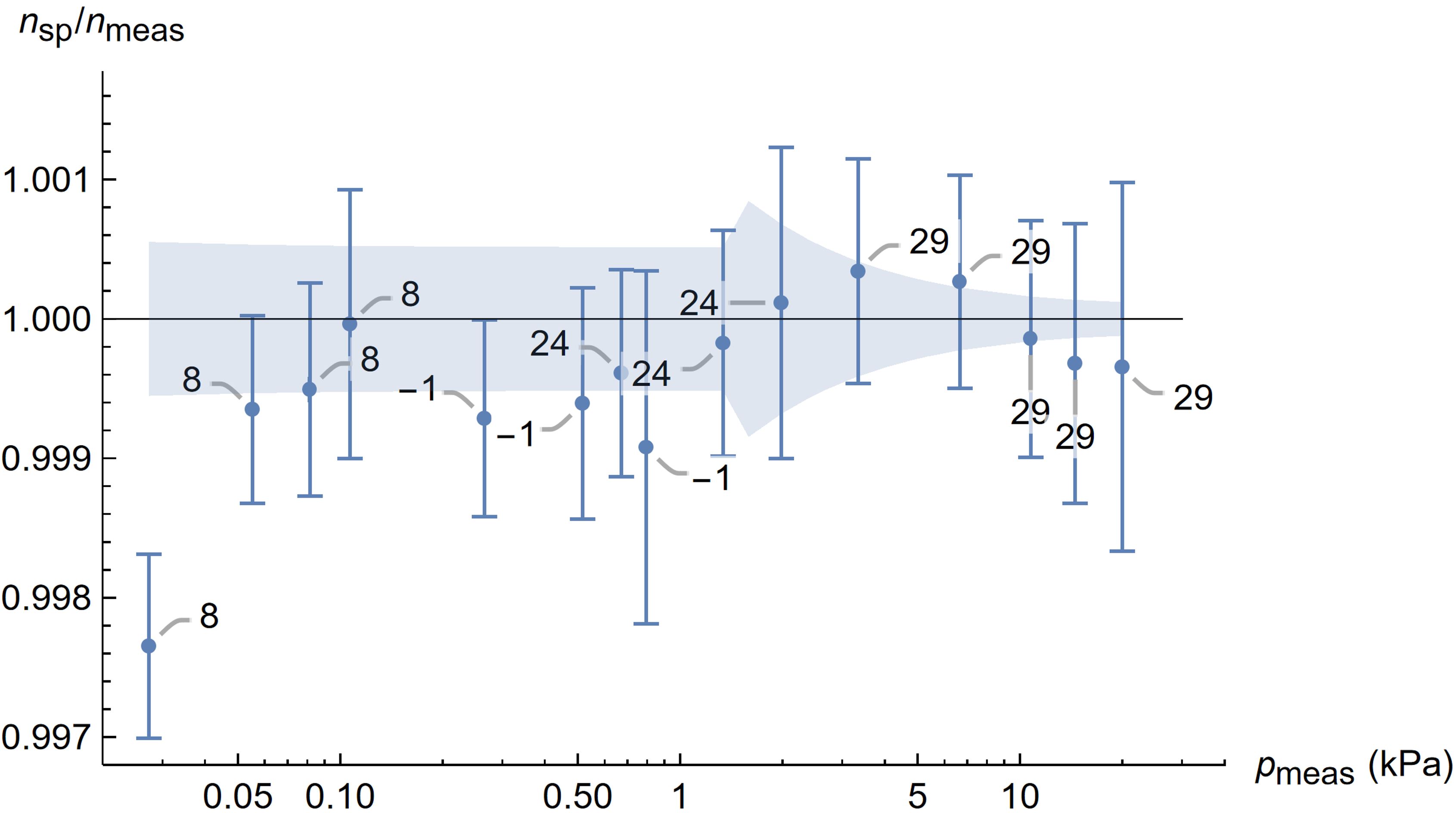}
\caption{Ratios of gas concentrations, $n$, obtained from spectroscopy and from the reference gas pressure and temperature measurements versus gas pressure $p_{\rm meas}$. The point labels indicate the spectral line $m$ index used for spectroscopic $n$ measurement. Error bars correspond to combined standard uncertainties. The gray area corresponds to the standard uncertainties of gas concentration measurements using calibrated $p$ and $T$ sensors.} \label{fig_nspec}
\end{figure}

Using Eq. \ref{Eqn} with $T$ determined from LRT, and the same experimental line areas, $A$, and {\it ab initio} line intensities, $S$, as earlier used to LRT, we determined absolute CO concentrations, $n$, in our samples. Here, we chose four lines with up to 240-fold different $S$ (between $9.39\times 10^{-26}$ and $2.26\times 10^{-23}$ cm/molecule) which enabled line shape measurement at CO pressures between 0.05 and 20 kPa. The comparison between spectroscopically determined concentration, $n_{\rm sp}$, and corresponding $n_{\rm meas}$ obtained from gas pressure measured with a pressure gauge calibrated to SI, using the equation of state of a real gas (with the second virial coefficient of $-8$ cm$^3$/mol \cite{Bousheheri1987}), is shown in Fig. \ref{fig_nspec}. The weighted mean ratio $n_{\rm sp}/n_{\rm meas}$ and its standard uncertainty obtained from the pressure range of 50 Pa to 20 kPa are $(0.99971 \pm 0.00025)$ while the standard deviation for a single pressure, corresponding to a single spectral line is 0.00038. The lowest-pressure point was excluded from the mean as an outlier. It shows the limitation of our pressure gauge linearity at a low-pressure range, as shown in Appendix \ref{app2}. 

\subsection{Uncertainty analysis}

The uncertainty components given in Table \ref{Tab_Uncert} were considered to estimate the combined uncertainties of our spectroscopic temperature and concentration measurements presented in Figs. \ref{fig_Tspec} and \ref{fig_nspec}. Note that the reference $T$ and $p$ measurements do not contribute to the spectroscopic measurements, but only to their comparison with traditional contact sensors. The line area fit uncertainty was estimated by comparing line areas obtained by fitting the Hartmann-Tran and the speed-dependent Voigt line shape models and including or neglecting residual etalon fringes barely visible in the spectra. The isotopic abundance of $^{12}$C$^{16}$O in the sample, contributing to uncertainty of $n$, was measured spectroscopically using line intensities of $^{12}$C$^{16}$O and $^{13}$C$^{16}$O \cite{CCQMP229}. The uncertainties of the reference {\it ab initio}  $S$ were assumed to be the difference between the calculated and multi-laboratory experimental values averaged over all $m$ reported in Ref. \cite{CCQMP229}. However, for P20 and P22 lines, uncertainties were increased to 1$\permil$, consistent with their intensity differences between theory and experiment in Ref. \cite{CCQMP229}. Results in Fig. \ref{fig_Tspec} also suggest decreased accuracy of ab initio $S$  for P20 and P22 lines. Finally, the uncertainty of $\Delta E''$ has a completely negligible impact on $u(T)$.
\begin{table}[!ht]
\caption{Uncertainty components ($1\sigma$) of spectroscopic gas temperature, $T$, and concentration, $n$, measurements. The values (range or average over line pairs) correspond to $T$ obtained from the two-line ratio and $n$ obtained from one line, except for the "combined (all lines)", which corresponds to the weighted mean from all used lines.}
\label{Tab_Uncert}
\begin{ruledtabular} 
\begin{tabular}{ccc}
uncertainty source & $u(T)/T$ ($\permil$) & $u(n)/n$ ($\permil$) \\
\hline
line area fit & 0.17 & 0.6  \\
$A$ vs $p$ linearity & 0.082 & --  \\
reference $S$ & 0.04 - 0.28 & 0.14 - 1 \\
mean $T$ from LRT & -- & 0.082   \\
$^{12}$C$^{16}$O abundance & -- & 0.13  \\
\hline
combined & 0.019 - 0.034 & 0.064 - 1.18 \\
{\bf combined (all lines)} & {\bf 0.082} & {\bf 0.250} \\
\hline
reference $T$ \footnote{For comparison with reference sensors only.} & 0.11 & 0.11  \\
reference $p$ \footnotemark[1] & -- & 0.12 - 0.7   \\
\end{tabular}
\end{ruledtabular} 
\end{table}

\subsection{Conclusions}

The introduced the LRT primary thermometry exploiting resonant frequencies of the optical cavity. The achieved uncertainty of our LRT, $u(T)/T$ of 82 ppm, is over six-fold lower than the best previously reported result \cite{Gotti2020} and covers three orders of magnitude in gas pressure - from tens of pascals to 20 kPa, without noticeable impact on accuracy. In the pressures above 1.2 kPa, our results are over an order of magnitude more accurate than any previously reported spectroscopic thermometry. This broad range of gas pressure significantly extends the applicability of LRT in primary non-contact thermometry, including its use in the optical amount of substance and pressure measurements. The LRT uncertainty budget shows that further accuracy improvement may be achieved mainly through improved accuracy of the reference {\it ab initio} line intensities. In the multi-laboratory comparison \cite{Bielska2022, CCQMP229}, {\it ab initio} $S$ show permille-level, $m$-dependent deviations from the experimental ones, which is consistent with our LRT using P- and R-branch lines. Also, the experimental line area can be further improved by minimizing residual etalon fringes in spectra and careful study of the collisional line-shape model.

In the case of fully spectroscopic, line intensity-based gas concentration measurement, we demonstrated the most accurate, to the best of our knowledge, result with the mean relative standard uncertainty, $u(n)/n$ of 250 ppm. Similarly to the spectroscopic $T$, further improvement of $n$ measurement is possible by improving the {\it ab initio} $S$ accuracy, and the spectrum model. Currently, the presented verification of spectroscopic $n$ has reached the accuracy limitation of our reference pressure sensors.  

The presented results demonstrate the potential of the spectroscopy of a simple CO molecule combined with fully frequency-based CMDS and {\it ab initio} line intensities to non-contact optical primary thermometry and amount-of-substance (per volume) metrology. The achieved uncertainty of the calibration-free spectroscopic $T$ and $n$ is already comparable to commercially available contact thermometers applied to gases and pressure gauges, which require regular calibrations. These results also show the opportunity for further improvement of the reference line intensities, through a joint effort of the experimental and theoretical groups \cite{Bielska2022, CCQMP229}. We note that the applicability of the presented methods for higher pressures would require careful collisional line-shape effects analysis, including line-mixing \cite{Ciurylo2001LM, Pine2019} and a breakdown of the impact approximation \cite{Boulet2004, Reed2023}.

\subsection*{Acknowledgments}

The research was supported by the National Science Centre, Poland, project Nos. 2023/51/B/ST2/00427 (D.L.), 2021/42/E/ST2/00152 (S.W.), 2020/39/B/ST2/00719 (A.C.), 2022/46/E/ST2/00282 (P.W.). 
The research was supported by the state budget of Poland, allocated by the Minister of Education and Science under the "Polska Metrologia II" program, project no. PM-II/SP/0011/2024/02, subsidy amount 988 900 PLN.
DL acknowledges partial funding support from the PriSpecTemp project. The project (22IEM03 PriSpecTemp) has received funding from the European Partnership on Metrology, co-financed from the European Union’s Horizon Europe Research and Innovation Programme and by the Participating States.


\begin{thebibliography}{50}%
\makeatletter
\providecommand \@ifxundefined [1]{%
 \@ifx{#1\undefined}
}%
\providecommand \@ifnum [1]{%
 \ifnum #1\expandafter \@firstoftwo
 \else \expandafter \@secondoftwo
 \fi
}%
\providecommand \@ifx [1]{%
 \ifx #1\expandafter \@firstoftwo
 \else \expandafter \@secondoftwo
 \fi
}%
\providecommand \natexlab [1]{#1}%
\providecommand \enquote  [1]{``#1''}%
\providecommand \bibnamefont  [1]{#1}%
\providecommand \bibfnamefont [1]{#1}%
\providecommand \citenamefont [1]{#1}%
\providecommand \href@noop [0]{\@secondoftwo}%
\providecommand \href [0]{\begingroup \@sanitize@url \@href}%
\providecommand \@href[1]{\@@startlink{#1}\@@href}%
\providecommand \@@href[1]{\endgroup#1\@@endlink}%
\providecommand \@sanitize@url [0]{\catcode `\\12\catcode `\$12\catcode `\&12\catcode `\#12\catcode `\^12\catcode `\_12\catcode `\%12\relax}%
\providecommand \@@startlink[1]{}%
\providecommand \@@endlink[0]{}%
\providecommand \url  [0]{\begingroup\@sanitize@url \@url }%
\providecommand \@url [1]{\endgroup\@href {#1}{\urlprefix }}%
\providecommand \urlprefix  [0]{URL }%
\providecommand \Eprint [0]{\href }%
\providecommand \doibase [0]{http://dx.doi.org/}%
\providecommand \selectlanguage [0]{\@gobble}%
\providecommand \bibinfo  [0]{\@secondoftwo}%
\providecommand \bibfield  [0]{\@secondoftwo}%
\providecommand \translation [1]{[#1]}%
\providecommand \BibitemOpen [0]{}%
\providecommand \bibitemStop [0]{}%
\providecommand \bibitemNoStop [0]{.\EOS\space}%
\providecommand \EOS [0]{\spacefactor3000\relax}%
\providecommand \BibitemShut  [1]{\csname bibitem#1\endcsname}%
\let\auto@bib@innerbib\@empty
\bibitem [{\citenamefont {Stock}\ \emph {et~al.}(2019)\citenamefont {Stock}, \citenamefont {Davis}, \citenamefont {de~Mirandes},\ and\ \citenamefont {Milton}}]{Stock2019}%
  \BibitemOpen
  \bibfield  {author} {\bibinfo {author} {\bibfnamefont {M.}~\bibnamefont {Stock}}, \bibinfo {author} {\bibfnamefont {R.}~\bibnamefont {Davis}}, \bibinfo {author} {\bibfnamefont {E.}~\bibnamefont {de~Mirandes}}, \ and\ \bibinfo {author} {\bibfnamefont {M.~J.~T.}\ \bibnamefont {Milton}},\ }\href {\doibase 10.1088/1681-7575/ab0013} {\bibfield  {journal} {\bibinfo  {journal} {Metrologia}\ }\textbf {\bibinfo {volume} {56}},\ \bibinfo {pages} {022001} (\bibinfo {year} {2019})}\BibitemShut {NoStop}%
\bibitem [{\citenamefont {Fellmuth}\ \emph {et~al.}(2016)\citenamefont {Fellmuth}, \citenamefont {Fischer}, \citenamefont {Machin}, \citenamefont {Picard}, \citenamefont {Steur}, \citenamefont {Tamura}, \citenamefont {White},\ and\ \citenamefont {Yoon}}]{Fellmuth2016}%
  \BibitemOpen
  \bibfield  {author} {\bibinfo {author} {\bibfnamefont {B.}~\bibnamefont {Fellmuth}}, \bibinfo {author} {\bibfnamefont {J.}~\bibnamefont {Fischer}}, \bibinfo {author} {\bibfnamefont {G.}~\bibnamefont {Machin}}, \bibinfo {author} {\bibfnamefont {S.}~\bibnamefont {Picard}}, \bibinfo {author} {\bibfnamefont {P.~P.~M.}\ \bibnamefont {Steur}}, \bibinfo {author} {\bibfnamefont {O.}~\bibnamefont {Tamura}}, \bibinfo {author} {\bibfnamefont {D.~R.}\ \bibnamefont {White}}, \ and\ \bibinfo {author} {\bibfnamefont {H.}~\bibnamefont {Yoon}},\ }\href {\doibase 10.1098/rsta.2015.0037} {\bibfield  {journal} {\bibinfo  {journal} {Phil. Trans. R. Soc. A}\ }\textbf {\bibinfo {volume} {374}},\ \bibinfo {pages} {20150037} (\bibinfo {year} {2016})}\BibitemShut {NoStop}%
\bibitem [{\citenamefont {Daussy}\ \emph {et~al.}(2007)\citenamefont {Daussy}, \citenamefont {Guinet}, \citenamefont {Amy-Klein}, \citenamefont {Djerroud}, \citenamefont {Hermier}, \citenamefont {Briaudeau}, \citenamefont {Bord\'e},\ and\ \citenamefont {Chardonnet}}]{Daussy2007}%
  \BibitemOpen
  \bibfield  {author} {\bibinfo {author} {\bibfnamefont {C.}~\bibnamefont {Daussy}}, \bibinfo {author} {\bibfnamefont {M.}~\bibnamefont {Guinet}}, \bibinfo {author} {\bibfnamefont {A.}~\bibnamefont {Amy-Klein}}, \bibinfo {author} {\bibfnamefont {K.}~\bibnamefont {Djerroud}}, \bibinfo {author} {\bibfnamefont {Y.}~\bibnamefont {Hermier}}, \bibinfo {author} {\bibfnamefont {S.}~\bibnamefont {Briaudeau}}, \bibinfo {author} {\bibfnamefont {C.~J.}\ \bibnamefont {Bord\'e}}, \ and\ \bibinfo {author} {\bibfnamefont {C.}~\bibnamefont {Chardonnet}},\ }\href {\doibase 10.1103/PhysRevLett.98.250801} {\bibfield  {journal} {\bibinfo  {journal} {Phys. Rev. Lett.}\ }\textbf {\bibinfo {volume} {98}},\ \bibinfo {pages} {250801} (\bibinfo {year} {2007})}\BibitemShut {NoStop}%
\bibitem [{\citenamefont {Gianfrani}(2016)}]{Gianfrani2016}%
  \BibitemOpen
  \bibfield  {author} {\bibinfo {author} {\bibfnamefont {L.}~\bibnamefont {Gianfrani}},\ }\href {\doibase 10.1098/rsta.2015.0047} {\bibfield  {journal} {\bibinfo  {journal} {Phil. Trans. R. Soc. A}\ }\textbf {\bibinfo {volume} {374}},\ \bibinfo {pages} {20150047} (\bibinfo {year} {2016})}\BibitemShut {NoStop}%
\bibitem [{\citenamefont {Arroyo}\ and\ \citenamefont {Hanson}(1993)}]{Arroyo1993}%
  \BibitemOpen
  \bibfield  {author} {\bibinfo {author} {\bibfnamefont {M.~P.}\ \bibnamefont {Arroyo}}\ and\ \bibinfo {author} {\bibfnamefont {R.~K.}\ \bibnamefont {Hanson}},\ }\href {\doibase 10.1364/AO.32.006104} {\bibfield  {journal} {\bibinfo  {journal} {Appl. Opt.}\ }\textbf {\bibinfo {volume} {32}},\ \bibinfo {pages} {6104} (\bibinfo {year} {1993})}\BibitemShut {NoStop}%
\bibitem [{\citenamefont {Berman}\ \emph {et~al.}(1993)\citenamefont {Berman}, \citenamefont {Duggan}, \citenamefont {Flohic}, \citenamefont {May},\ and\ \citenamefont {Drummond}}]{Berman1993}%
  \BibitemOpen
  \bibfield  {author} {\bibinfo {author} {\bibfnamefont {R.}~\bibnamefont {Berman}}, \bibinfo {author} {\bibfnamefont {P.}~\bibnamefont {Duggan}}, \bibinfo {author} {\bibfnamefont {M.~P.~L.}\ \bibnamefont {Flohic}}, \bibinfo {author} {\bibfnamefont {A.~D.}\ \bibnamefont {May}}, \ and\ \bibinfo {author} {\bibfnamefont {J.~R.}\ \bibnamefont {Drummond}},\ }\href {\doibase 10.1364/AO.32.006280} {\bibfield  {journal} {\bibinfo  {journal} {Appl. Opt.}\ }\textbf {\bibinfo {volume} {32}},\ \bibinfo {pages} {6280} (\bibinfo {year} {1993})}\BibitemShut {NoStop}%
\bibitem [{\citenamefont {Shimizu}\ \emph {et~al.}(2018)\citenamefont {Shimizu}, \citenamefont {Okubo}, \citenamefont {Onae}, \citenamefont {Yamada},\ and\ \citenamefont {Inaba}}]{Shimizu2018}%
  \BibitemOpen
  \bibfield  {author} {\bibinfo {author} {\bibfnamefont {Y.}~\bibnamefont {Shimizu}}, \bibinfo {author} {\bibfnamefont {S.}~\bibnamefont {Okubo}}, \bibinfo {author} {\bibfnamefont {A.}~\bibnamefont {Onae}}, \bibinfo {author} {\bibfnamefont {K.~M.~T.}\ \bibnamefont {Yamada}}, \ and\ \bibinfo {author} {\bibfnamefont {H.}~\bibnamefont {Inaba}},\ }\href {\doibase 10.1007/s00340-018-6933-x} {\bibfield  {journal} {\bibinfo  {journal} {Appl. Phys. B}\ }\textbf {\bibinfo {volume} {124}},\ \bibinfo {pages} {71} (\bibinfo {year} {2018})}\BibitemShut {NoStop}%
\bibitem [{\citenamefont {Santamaria~Amato}\ \emph {et~al.}(2019)\citenamefont {Santamaria~Amato}, \citenamefont {de~Cumis}, \citenamefont {Bianco}, \citenamefont {Pastore},\ and\ \citenamefont {Pastor}}]{SantamariaAmato2019}%
  \BibitemOpen
  \bibfield  {author} {\bibinfo {author} {\bibfnamefont {L.}~\bibnamefont {Santamaria~Amato}}, \bibinfo {author} {\bibfnamefont {M.~S.}\ \bibnamefont {de~Cumis}}, \bibinfo {author} {\bibfnamefont {G.}~\bibnamefont {Bianco}}, \bibinfo {author} {\bibfnamefont {R.}~\bibnamefont {Pastore}}, \ and\ \bibinfo {author} {\bibfnamefont {P.~C.}\ \bibnamefont {Pastor}},\ }\href {\doibase 10.1088/1367-2630/ab4d07} {\bibfield  {journal} {\bibinfo  {journal} {New J. Phys.}\ }\textbf {\bibinfo {volume} {21}},\ \bibinfo {pages} {113008} (\bibinfo {year} {2019})}\BibitemShut {NoStop}%
\bibitem [{\citenamefont {Gotti}\ \emph {et~al.}(2020)\citenamefont {Gotti}, \citenamefont {Lamperti}, \citenamefont {Gatti}, \citenamefont {Wójtewicz}, \citenamefont {Puppe}, \citenamefont {Mayzlin}, \citenamefont {Alsaif}, \citenamefont {Robinson-Tait}, \citenamefont {Rohde}, \citenamefont {Wilk}, \citenamefont {Leisching}, \citenamefont {Kaenders}, \citenamefont {Laporta},\ and\ \citenamefont {Marangoni}}]{Gotti2020}%
  \BibitemOpen
  \bibfield  {author} {\bibinfo {author} {\bibfnamefont {R.}~\bibnamefont {Gotti}}, \bibinfo {author} {\bibfnamefont {M.}~\bibnamefont {Lamperti}}, \bibinfo {author} {\bibfnamefont {D.}~\bibnamefont {Gatti}}, \bibinfo {author} {\bibfnamefont {S.}~\bibnamefont {Wójtewicz}}, \bibinfo {author} {\bibfnamefont {T.}~\bibnamefont {Puppe}}, \bibinfo {author} {\bibfnamefont {Y.}~\bibnamefont {Mayzlin}}, \bibinfo {author} {\bibfnamefont {B.}~\bibnamefont {Alsaif}}, \bibinfo {author} {\bibfnamefont {J.}~\bibnamefont {Robinson-Tait}}, \bibinfo {author} {\bibfnamefont {F.}~\bibnamefont {Rohde}}, \bibinfo {author} {\bibfnamefont {R.}~\bibnamefont {Wilk}}, \bibinfo {author} {\bibfnamefont {P.}~\bibnamefont {Leisching}}, \bibinfo {author} {\bibfnamefont {W.~G.}\ \bibnamefont {Kaenders}}, \bibinfo {author} {\bibfnamefont {P.}~\bibnamefont {Laporta}}, \ and\ \bibinfo {author} {\bibfnamefont {M.}~\bibnamefont {Marangoni}},\ }\href {\doibase 10.1088/1367-2630/aba85d} {\bibfield  {journal} {\bibinfo  {journal} {New J. Phys.}\
  }\textbf {\bibinfo {volume} {22}},\ \bibinfo {pages} {083071} (\bibinfo {year} {2020})}\BibitemShut {NoStop}%
\bibitem [{\citenamefont {Ciury\l{}o}\ \emph {et~al.}(2001)\citenamefont {Ciury\l{}o}, \citenamefont {Pine},\ and\ \citenamefont {Szudy}}]{Ciurylo2001JQSRT}%
  \BibitemOpen
  \bibfield  {author} {\bibinfo {author} {\bibfnamefont {R.}~\bibnamefont {Ciury\l{}o}}, \bibinfo {author} {\bibfnamefont {A.}~\bibnamefont {Pine}}, \ and\ \bibinfo {author} {\bibfnamefont {J.}~\bibnamefont {Szudy}},\ }\href {\doibase 10.1016/S0022-4073(00)00024-8} {\bibfield  {journal} {\bibinfo  {journal} {J. Quant. Spectrosc. Radiat. Transfer}\ }\textbf {\bibinfo {volume} {68}},\ \bibinfo {pages} {257 } (\bibinfo {year} {2001})}\BibitemShut {NoStop}%
\bibitem [{\citenamefont {Ciury\l{}o}\ \emph {et~al.}(2002)\citenamefont {Ciury\l{}o}, \citenamefont {Shapiro}, \citenamefont {Drummond},\ and\ \citenamefont {May}}]{Ciurylo2002BB}%
  \BibitemOpen
  \bibfield  {author} {\bibinfo {author} {\bibfnamefont {R.}~\bibnamefont {Ciury\l{}o}}, \bibinfo {author} {\bibfnamefont {D.~A.}\ \bibnamefont {Shapiro}}, \bibinfo {author} {\bibfnamefont {J.~R.}\ \bibnamefont {Drummond}}, \ and\ \bibinfo {author} {\bibfnamefont {A.~D.}\ \bibnamefont {May}},\ }\href {\doibase 10.1103/PhysRevA.65.012502} {\bibfield  {journal} {\bibinfo  {journal} {Phys. Rev. A}\ }\textbf {\bibinfo {volume} {65}},\ \bibinfo {pages} {012502} (\bibinfo {year} {2002})}\BibitemShut {NoStop}%
\bibitem [{\citenamefont {Ngo}\ \emph {et~al.}(2013)\citenamefont {Ngo}, \citenamefont {Lisak}, \citenamefont {Tran},\ and\ \citenamefont {Hartmann}}]{Ngo2013}%
  \BibitemOpen
  \bibfield  {author} {\bibinfo {author} {\bibfnamefont {N.}~\bibnamefont {Ngo}}, \bibinfo {author} {\bibfnamefont {D.}~\bibnamefont {Lisak}}, \bibinfo {author} {\bibfnamefont {H.}~\bibnamefont {Tran}}, \ and\ \bibinfo {author} {\bibfnamefont {J.-M.}\ \bibnamefont {Hartmann}},\ }\href {\doibase 10.1016/j.jqsrt.2013.05.034} {\bibfield  {journal} {\bibinfo  {journal} {J. Quant. Spectrosc. Radiat. Transf.}\ }\textbf {\bibinfo {volume} {129}},\ \bibinfo {pages} {89} (\bibinfo {year} {2013})}\BibitemShut {NoStop}%
\bibitem [{\citenamefont {Cygan}\ \emph {et~al.}(2010)\citenamefont {Cygan}, \citenamefont {Lisak}, \citenamefont {Trawi\ifmmode~\acute{n}\else \'{n}\fi{}ski},\ and\ \citenamefont {Ciury\l{}o}}]{Cygan2010}%
  \BibitemOpen
  \bibfield  {author} {\bibinfo {author} {\bibfnamefont {A.}~\bibnamefont {Cygan}}, \bibinfo {author} {\bibfnamefont {D.}~\bibnamefont {Lisak}}, \bibinfo {author} {\bibfnamefont {R.~S.}\ \bibnamefont {Trawi\ifmmode~\acute{n}\else \'{n}\fi{}ski}}, \ and\ \bibinfo {author} {\bibfnamefont {R.}~\bibnamefont {Ciury\l{}o}},\ }\href {\doibase 10.1103/PhysRevA.82.032515} {\bibfield  {journal} {\bibinfo  {journal} {Phys. Rev. A}\ }\textbf {\bibinfo {volume} {82}},\ \bibinfo {pages} {032515} (\bibinfo {year} {2010})}\BibitemShut {NoStop}%
\bibitem [{\citenamefont {Triki}\ \emph {et~al.}(2012)\citenamefont {Triki}, \citenamefont {Lemarchand}, \citenamefont {Darqui\'e}, \citenamefont {Sow}, \citenamefont {Roncin}, \citenamefont {Chardonnet},\ and\ \citenamefont {Daussy}}]{Triki2012}%
  \BibitemOpen
  \bibfield  {author} {\bibinfo {author} {\bibfnamefont {M.}~\bibnamefont {Triki}}, \bibinfo {author} {\bibfnamefont {C.}~\bibnamefont {Lemarchand}}, \bibinfo {author} {\bibfnamefont {B.}~\bibnamefont {Darqui\'e}}, \bibinfo {author} {\bibfnamefont {P.~L.~T.}\ \bibnamefont {Sow}}, \bibinfo {author} {\bibfnamefont {V.}~\bibnamefont {Roncin}}, \bibinfo {author} {\bibfnamefont {C.}~\bibnamefont {Chardonnet}}, \ and\ \bibinfo {author} {\bibfnamefont {C.}~\bibnamefont {Daussy}},\ }\href {\doibase 10.1103/PhysRevA.85.062510} {\bibfield  {journal} {\bibinfo  {journal} {Phys. Rev. A}\ }\textbf {\bibinfo {volume} {85}},\ \bibinfo {pages} {062510} (\bibinfo {year} {2012})}\BibitemShut {NoStop}%
\bibitem [{\citenamefont {Moretti}\ \emph {et~al.}(2013)\citenamefont {Moretti}, \citenamefont {Castrillo}, \citenamefont {Fasci}, \citenamefont {De~Vizia}, \citenamefont {Casa}, \citenamefont {Galzerano}, \citenamefont {Merlone}, \citenamefont {Laporta},\ and\ \citenamefont {Gianfrani}}]{Moretti2013}%
  \BibitemOpen
  \bibfield  {author} {\bibinfo {author} {\bibfnamefont {L.}~\bibnamefont {Moretti}}, \bibinfo {author} {\bibfnamefont {A.}~\bibnamefont {Castrillo}}, \bibinfo {author} {\bibfnamefont {E.}~\bibnamefont {Fasci}}, \bibinfo {author} {\bibfnamefont {M.~D.}\ \bibnamefont {De~Vizia}}, \bibinfo {author} {\bibfnamefont {G.}~\bibnamefont {Casa}}, \bibinfo {author} {\bibfnamefont {G.}~\bibnamefont {Galzerano}}, \bibinfo {author} {\bibfnamefont {A.}~\bibnamefont {Merlone}}, \bibinfo {author} {\bibfnamefont {P.}~\bibnamefont {Laporta}}, \ and\ \bibinfo {author} {\bibfnamefont {L.}~\bibnamefont {Gianfrani}},\ }\href {\doibase 10.1103/PhysRevLett.111.060803} {\bibfield  {journal} {\bibinfo  {journal} {Phys. Rev. Lett.}\ }\textbf {\bibinfo {volume} {111}},\ \bibinfo {pages} {060803} (\bibinfo {year} {2013})}\BibitemShut {NoStop}%
\bibitem [{\citenamefont {Wcis\l{}o}\ \emph {et~al.}(2013)\citenamefont {Wcis\l{}o}, \citenamefont {Cygan}, \citenamefont {Lisak},\ and\ \citenamefont {Ciury\l{}o}}]{Wcislo2013}%
  \BibitemOpen
  \bibfield  {author} {\bibinfo {author} {\bibfnamefont {P.}~\bibnamefont {Wcis\l{}o}}, \bibinfo {author} {\bibfnamefont {A.}~\bibnamefont {Cygan}}, \bibinfo {author} {\bibfnamefont {D.}~\bibnamefont {Lisak}}, \ and\ \bibinfo {author} {\bibfnamefont {R.}~\bibnamefont {Ciury\l{}o}},\ }\href {\doibase 10.1103/PhysRevA.88.012517} {\bibfield  {journal} {\bibinfo  {journal} {Phys. Rev. A}\ }\textbf {\bibinfo {volume} {88}},\ \bibinfo {pages} {012517} (\bibinfo {year} {2013})}\BibitemShut {NoStop}%
\bibitem [{\citenamefont {Rothman}\ \emph {et~al.}(1998)\citenamefont {Rothman}, \citenamefont {Rinsland}, \citenamefont {Goldman}, \citenamefont {Massie}, \citenamefont {Edwards}, \citenamefont {Flaud}, \citenamefont {Perrin}, \citenamefont {Camy-Peyret}, \citenamefont {Dana}, \citenamefont {Mandin}, \citenamefont {Schroeder}, \citenamefont {Mccann}, \citenamefont {Gamache}, \citenamefont {Wattson}, \citenamefont {Yoshino}, \citenamefont {Chance}, \citenamefont {Jucks}, \citenamefont {Brown}, \citenamefont {Nemtchinov},\ and\ \citenamefont {Varanasi}}]{Rothman1996}%
  \BibitemOpen
  \bibfield  {author} {\bibinfo {author} {\bibfnamefont {L.}~\bibnamefont {Rothman}}, \bibinfo {author} {\bibfnamefont {C.}~\bibnamefont {Rinsland}}, \bibinfo {author} {\bibfnamefont {A.}~\bibnamefont {Goldman}}, \bibinfo {author} {\bibfnamefont {S.}~\bibnamefont {Massie}}, \bibinfo {author} {\bibfnamefont {D.}~\bibnamefont {Edwards}}, \bibinfo {author} {\bibfnamefont {J.-M.}\ \bibnamefont {Flaud}}, \bibinfo {author} {\bibfnamefont {A.}~\bibnamefont {Perrin}}, \bibinfo {author} {\bibfnamefont {C.}~\bibnamefont {Camy-Peyret}}, \bibinfo {author} {\bibfnamefont {V.}~\bibnamefont {Dana}}, \bibinfo {author} {\bibfnamefont {J.-Y.}\ \bibnamefont {Mandin}}, \bibinfo {author} {\bibfnamefont {J.}~\bibnamefont {Schroeder}}, \bibinfo {author} {\bibfnamefont {A.}~\bibnamefont {Mccann}}, \bibinfo {author} {\bibfnamefont {R.}~\bibnamefont {Gamache}}, \bibinfo {author} {\bibfnamefont {R.}~\bibnamefont {Wattson}}, \bibinfo {author} {\bibfnamefont {K.}~\bibnamefont {Yoshino}}, \bibinfo {author} {\bibfnamefont {K.}~\bibnamefont
  {Chance}}, \bibinfo {author} {\bibfnamefont {K.}~\bibnamefont {Jucks}}, \bibinfo {author} {\bibfnamefont {L.}~\bibnamefont {Brown}}, \bibinfo {author} {\bibfnamefont {V.}~\bibnamefont {Nemtchinov}}, \ and\ \bibinfo {author} {\bibfnamefont {P.}~\bibnamefont {Varanasi}},\ }\href {\doibase 10.1016/S0022-4073(98)00078-8} {\bibfield  {journal} {\bibinfo  {journal} {J. Quant. Spectrosc. Radiat. Transf.}\ }\textbf {\bibinfo {volume} {60}},\ \bibinfo {pages} {665} (\bibinfo {year} {1998})}\BibitemShut {NoStop}%
\bibitem [{\citenamefont {Santamaria~Amato}\ \emph {et~al.}(2018)\citenamefont {Santamaria~Amato}, \citenamefont {Siciliani~de Cumis}, \citenamefont {Dequal}, \citenamefont {Bianco},\ and\ \citenamefont {Pablo}}]{SantamariaAmato2018}%
  \BibitemOpen
  \bibfield  {author} {\bibinfo {author} {\bibfnamefont {L.}~\bibnamefont {Santamaria~Amato}}, \bibinfo {author} {\bibfnamefont {M.}~\bibnamefont {Siciliani~de Cumis}}, \bibinfo {author} {\bibfnamefont {D.}~\bibnamefont {Dequal}}, \bibinfo {author} {\bibfnamefont {G.}~\bibnamefont {Bianco}}, \ and\ \bibinfo {author} {\bibfnamefont {C.~P.}\ \bibnamefont {Pablo}},\ }\href {\doibase 10.1021/acs.jpca.8b05523} {\bibfield  {journal} {\bibinfo  {journal} {J. Phys. Chem. A}\ }\textbf {\bibinfo {volume} {122}},\ \bibinfo {pages} {6026} (\bibinfo {year} {2018})}\BibitemShut {NoStop}%
\bibitem [{\citenamefont {Pachucki}\ and\ \citenamefont {Komasa}(2008)}]{Pachucki2008}%
  \BibitemOpen
  \bibfield  {author} {\bibinfo {author} {\bibfnamefont {K.}~\bibnamefont {Pachucki}}\ and\ \bibinfo {author} {\bibfnamefont {J.}~\bibnamefont {Komasa}},\ }\href {\doibase 10.1103/PhysRevA.78.052503} {\bibfield  {journal} {\bibinfo  {journal} {Phys. Rev. A}\ }\textbf {\bibinfo {volume} {78}},\ \bibinfo {pages} {052503} (\bibinfo {year} {2008})}\BibitemShut {NoStop}%
\bibitem [{\citenamefont {Polyansky}\ \emph {et~al.}(2015)\citenamefont {Polyansky}, \citenamefont {Bielska}, \citenamefont {Ghysels}, \citenamefont {Lodi}, \citenamefont {Zobov}, \citenamefont {Hodges},\ and\ \citenamefont {Tennyson}}]{Polyansky2015}%
  \BibitemOpen
  \bibfield  {author} {\bibinfo {author} {\bibfnamefont {O.~L.}\ \bibnamefont {Polyansky}}, \bibinfo {author} {\bibfnamefont {K.}~\bibnamefont {Bielska}}, \bibinfo {author} {\bibfnamefont {M.}~\bibnamefont {Ghysels}}, \bibinfo {author} {\bibfnamefont {L.}~\bibnamefont {Lodi}}, \bibinfo {author} {\bibfnamefont {N.~F.}\ \bibnamefont {Zobov}}, \bibinfo {author} {\bibfnamefont {J.~T.}\ \bibnamefont {Hodges}}, \ and\ \bibinfo {author} {\bibfnamefont {J.}~\bibnamefont {Tennyson}},\ }\href {\doibase 10.1103/PhysRevLett.114.243001} {\bibfield  {journal} {\bibinfo  {journal} {Phys. Rev. Lett.}\ }\textbf {\bibinfo {volume} {114}},\ \bibinfo {pages} {243001} (\bibinfo {year} {2015})}\BibitemShut {NoStop}%
\bibitem [{\citenamefont {Bielska}\ \emph {et~al.}(2022)\citenamefont {Bielska}, \citenamefont {Kyuberis}, \citenamefont {Reed}, \citenamefont {Li}, \citenamefont {Cygan}, \citenamefont {Ciury\l{}o}, \citenamefont {Adkins}, \citenamefont {Lodi}, \citenamefont {Zobov}, \citenamefont {Ebert}, \citenamefont {Lisak}, \citenamefont {Hodges}, \citenamefont {Tennyson},\ and\ \citenamefont {Polyansky}}]{Bielska2022}%
  \BibitemOpen
  \bibfield  {author} {\bibinfo {author} {\bibfnamefont {K.}~\bibnamefont {Bielska}}, \bibinfo {author} {\bibfnamefont {A.~A.}\ \bibnamefont {Kyuberis}}, \bibinfo {author} {\bibfnamefont {Z.~D.}\ \bibnamefont {Reed}}, \bibinfo {author} {\bibfnamefont {G.}~\bibnamefont {Li}}, \bibinfo {author} {\bibfnamefont {A.}~\bibnamefont {Cygan}}, \bibinfo {author} {\bibfnamefont {R.}~\bibnamefont {Ciury\l{}o}}, \bibinfo {author} {\bibfnamefont {E.~M.}\ \bibnamefont {Adkins}}, \bibinfo {author} {\bibfnamefont {L.}~\bibnamefont {Lodi}}, \bibinfo {author} {\bibfnamefont {N.~F.}\ \bibnamefont {Zobov}}, \bibinfo {author} {\bibfnamefont {V.}~\bibnamefont {Ebert}}, \bibinfo {author} {\bibfnamefont {D.}~\bibnamefont {Lisak}}, \bibinfo {author} {\bibfnamefont {J.~T.}\ \bibnamefont {Hodges}}, \bibinfo {author} {\bibfnamefont {J.}~\bibnamefont {Tennyson}}, \ and\ \bibinfo {author} {\bibfnamefont {O.~L.}\ \bibnamefont {Polyansky}},\ }\href {\doibase 10.1103/PhysRevLett.129.043002} {\bibfield  {journal} {\bibinfo  {journal} {Phys.
  Rev. Lett.}\ }\textbf {\bibinfo {volume} {129}},\ \bibinfo {pages} {043002} (\bibinfo {year} {2022})}\BibitemShut {NoStop}%
\bibitem [{\citenamefont {Boulet}\ \emph {et~al.}(2004)\citenamefont {Boulet}, \citenamefont {Flaud},\ and\ \citenamefont {Hartmann}}]{Boulet2004}%
  \BibitemOpen
  \bibfield  {author} {\bibinfo {author} {\bibfnamefont {C.}~\bibnamefont {Boulet}}, \bibinfo {author} {\bibfnamefont {P.-M.}\ \bibnamefont {Flaud}}, \ and\ \bibinfo {author} {\bibfnamefont {J.-M.}\ \bibnamefont {Hartmann}},\ }\href {\doibase 10.1063/1.1714794} {\bibfield  {journal} {\bibinfo  {journal} {The Journal of Chemical Physics}\ }\textbf {\bibinfo {volume} {120}},\ \bibinfo {pages} {11053} (\bibinfo {year} {2004})}\BibitemShut {NoStop}%
\bibitem [{\citenamefont {Reed}\ \emph {et~al.}(2023)\citenamefont {Reed}, \citenamefont {Tran}, \citenamefont {Ngo}, \citenamefont {Hartmann},\ and\ \citenamefont {Hodges}}]{Reed2023}%
  \BibitemOpen
  \bibfield  {author} {\bibinfo {author} {\bibfnamefont {Z.~D.}\ \bibnamefont {Reed}}, \bibinfo {author} {\bibfnamefont {H.}~\bibnamefont {Tran}}, \bibinfo {author} {\bibfnamefont {H.~N.}\ \bibnamefont {Ngo}}, \bibinfo {author} {\bibfnamefont {J.-M.}\ \bibnamefont {Hartmann}}, \ and\ \bibinfo {author} {\bibfnamefont {J.~T.}\ \bibnamefont {Hodges}},\ }\href {\doibase 10.1103/PhysRevLett.130.143001} {\bibfield  {journal} {\bibinfo  {journal} {Phys. Rev. Lett.}\ }\textbf {\bibinfo {volume} {130}},\ \bibinfo {pages} {143001} (\bibinfo {year} {2023})}\BibitemShut {NoStop}%
\bibitem [{\citenamefont {Ciury\l{}o}\ and\ \citenamefont {Szudy}(2001)}]{Ciurylo2001LM}%
  \BibitemOpen
  \bibfield  {author} {\bibinfo {author} {\bibfnamefont {R.}~\bibnamefont {Ciury\l{}o}}\ and\ \bibinfo {author} {\bibfnamefont {J.}~\bibnamefont {Szudy}},\ }\href {\doibase 10.1103/PhysRevA.63.042714} {\bibfield  {journal} {\bibinfo  {journal} {Phys. Rev. A}\ }\textbf {\bibinfo {volume} {63}},\ \bibinfo {pages} {042714} (\bibinfo {year} {2001})}\BibitemShut {NoStop}%
\bibitem [{\citenamefont {Pine}(2019)}]{Pine2019}%
  \BibitemOpen
  \bibfield  {author} {\bibinfo {author} {\bibfnamefont {A.}~\bibnamefont {Pine}},\ }\href {\doibase https://doi.org/10.1016/j.jqsrt.2018.10.038} {\bibfield  {journal} {\bibinfo  {journal} {Journal of Quantitative Spectroscopy and Radiative Transfer}\ }\textbf {\bibinfo {volume} {224}},\ \bibinfo {pages} {62} (\bibinfo {year} {2019})}\BibitemShut {NoStop}%
\bibitem [{\citenamefont {Romanini}\ \emph {et~al.}(1997)\citenamefont {Romanini}, \citenamefont {Kachanov}, \citenamefont {Sadeghi},\ and\ \citenamefont {Stoeckel}}]{Romanini1997}%
  \BibitemOpen
  \bibfield  {author} {\bibinfo {author} {\bibfnamefont {D.}~\bibnamefont {Romanini}}, \bibinfo {author} {\bibfnamefont {A.}~\bibnamefont {Kachanov}}, \bibinfo {author} {\bibfnamefont {N.}~\bibnamefont {Sadeghi}}, \ and\ \bibinfo {author} {\bibfnamefont {F.}~\bibnamefont {Stoeckel}},\ }\href {\doibase https://doi.org/10.1016/S0009-2614(96)01351-6} {\bibfield  {journal} {\bibinfo  {journal} {Chem. Phys. Lett.}\ }\textbf {\bibinfo {volume} {264}},\ \bibinfo {pages} {316} (\bibinfo {year} {1997})}\BibitemShut {NoStop}%
\bibitem [{\citenamefont {Cygan}\ \emph {et~al.}(2021)\citenamefont {Cygan}, \citenamefont {Fleisher}, \citenamefont {Ciuryło}, \citenamefont {Gillis}, \citenamefont {Hodges},\ and\ \citenamefont {Lisak}}]{Cygan2021}%
  \BibitemOpen
  \bibfield  {author} {\bibinfo {author} {\bibfnamefont {A.}~\bibnamefont {Cygan}}, \bibinfo {author} {\bibfnamefont {A.~J.}\ \bibnamefont {Fleisher}}, \bibinfo {author} {\bibfnamefont {R.}~\bibnamefont {Ciuryło}}, \bibinfo {author} {\bibfnamefont {K.~A.}\ \bibnamefont {Gillis}}, \bibinfo {author} {\bibfnamefont {J.~T.}\ \bibnamefont {Hodges}}, \ and\ \bibinfo {author} {\bibfnamefont {D.}~\bibnamefont {Lisak}},\ }\href {\doibase 10.1038/s42005-021-00517-3} {\bibfield  {journal} {\bibinfo  {journal} {Communic. Phys.}\ }\textbf {\bibinfo {volume} {4}},\ \bibinfo {pages} {14} (\bibinfo {year} {2021})}\BibitemShut {NoStop}%
\bibitem [{\citenamefont {Cygan}\ \emph {et~al.}(2013)\citenamefont {Cygan}, \citenamefont {Lisak}, \citenamefont {Morzy\'{n}ski}, \citenamefont {Bober}, \citenamefont {Zawada}, \citenamefont {Pazderski},\ and\ \citenamefont {Ciury{\l}o}}]{Cygan2013}%
  \BibitemOpen
  \bibfield  {author} {\bibinfo {author} {\bibfnamefont {A.}~\bibnamefont {Cygan}}, \bibinfo {author} {\bibfnamefont {D.}~\bibnamefont {Lisak}}, \bibinfo {author} {\bibfnamefont {P.}~\bibnamefont {Morzy\'{n}ski}}, \bibinfo {author} {\bibfnamefont {M.}~\bibnamefont {Bober}}, \bibinfo {author} {\bibfnamefont {M.}~\bibnamefont {Zawada}}, \bibinfo {author} {\bibfnamefont {E.}~\bibnamefont {Pazderski}}, \ and\ \bibinfo {author} {\bibfnamefont {R.}~\bibnamefont {Ciury{\l}o}},\ }\href {\doibase 10.1364/OE.21.029744} {\bibfield  {journal} {\bibinfo  {journal} {Opt. Express}\ }\textbf {\bibinfo {volume} {21}},\ \bibinfo {pages} {29744} (\bibinfo {year} {2013})}\BibitemShut {NoStop}%
\bibitem [{\citenamefont {Cygan}\ \emph {et~al.}(2015)\citenamefont {Cygan}, \citenamefont {Wcis{\l}o}, \citenamefont {W\'{o}jtewicz}, \citenamefont {Mas{\l}owski}, \citenamefont {Hodges}, \citenamefont {Ciury{\l}o},\ and\ \citenamefont {Lisak}}]{Cygan2015}%
  \BibitemOpen
  \bibfield  {author} {\bibinfo {author} {\bibfnamefont {A.}~\bibnamefont {Cygan}}, \bibinfo {author} {\bibfnamefont {P.}~\bibnamefont {Wcis{\l}o}}, \bibinfo {author} {\bibfnamefont {S.}~\bibnamefont {W\'{o}jtewicz}}, \bibinfo {author} {\bibfnamefont {P.}~\bibnamefont {Mas{\l}owski}}, \bibinfo {author} {\bibfnamefont {J.~T.}\ \bibnamefont {Hodges}}, \bibinfo {author} {\bibfnamefont {R.}~\bibnamefont {Ciury{\l}o}}, \ and\ \bibinfo {author} {\bibfnamefont {D.}~\bibnamefont {Lisak}},\ }\href {\doibase 10.1364/OE.23.014472} {\bibfield  {journal} {\bibinfo  {journal} {Opt. Express}\ }\textbf {\bibinfo {volume} {23}},\ \bibinfo {pages} {14472} (\bibinfo {year} {2015})}\BibitemShut {NoStop}%
\bibitem [{\citenamefont {Cygan}\ \emph {et~al.}(2025)\citenamefont {Cygan}, \citenamefont {W\'{o}jtewicz}, \citenamefont {J\'{o}\'{z}wiak}, \citenamefont {Kowzan}, \citenamefont {Stolarczyk}, \citenamefont {Bielska}, \citenamefont {Wcis{\l}o}, \citenamefont {Ciury{\l}o},\ and\ \citenamefont {Lisak}}]{Cygan2025}%
  \BibitemOpen
  \bibfield  {author} {\bibinfo {author} {\bibfnamefont {A.}~\bibnamefont {Cygan}}, \bibinfo {author} {\bibfnamefont {S.}~\bibnamefont {W\'{o}jtewicz}}, \bibinfo {author} {\bibfnamefont {H.}~\bibnamefont {J\'{o}\'{z}wiak}}, \bibinfo {author} {\bibfnamefont {G.}~\bibnamefont {Kowzan}}, \bibinfo {author} {\bibfnamefont {N.}~\bibnamefont {Stolarczyk}}, \bibinfo {author} {\bibfnamefont {K.}~\bibnamefont {Bielska}}, \bibinfo {author} {\bibfnamefont {P.}~\bibnamefont {Wcis{\l}o}}, \bibinfo {author} {\bibfnamefont {R.}~\bibnamefont {Ciury{\l}o}}, \ and\ \bibinfo {author} {\bibfnamefont {D.}~\bibnamefont {Lisak}},\ }\href {\doibase 10.1126/sciadv.adp8556} {\bibfield  {journal} {\bibinfo  {journal} {Sci. Adv.}\ }\textbf {\bibinfo {volume} {11}},\ \bibinfo {pages} {eadp8556} (\bibinfo {year} {2025})}\BibitemShut {NoStop}%
\bibitem [{\citenamefont {Gotti}\ \emph {et~al.}(2021)\citenamefont {Gotti}, \citenamefont {Lamperti}, \citenamefont {Gatti},\ and\ \citenamefont {Marangoni}}]{Gotti2021}%
  \BibitemOpen
  \bibfield  {author} {\bibinfo {author} {\bibfnamefont {R.}~\bibnamefont {Gotti}}, \bibinfo {author} {\bibfnamefont {M.}~\bibnamefont {Lamperti}}, \bibinfo {author} {\bibfnamefont {D.}~\bibnamefont {Gatti}}, \ and\ \bibinfo {author} {\bibfnamefont {M.}~\bibnamefont {Marangoni}},\ }\href {\doibase 10.1063/5.0055297} {\bibfield  {journal} {\bibinfo  {journal} {J. Phys. Chem. Ref. Data}\ }\textbf {\bibinfo {volume} {50}},\ \bibinfo {pages} {031501} (\bibinfo {year} {2021})}\BibitemShut {NoStop}%
\bibitem [{\citenamefont {Hashemi}\ \emph {et~al.}(2014)\citenamefont {Hashemi}, \citenamefont {Povey}, \citenamefont {Derksen}, \citenamefont {Naseri}, \citenamefont {Garber},\ and\ \citenamefont {Predoi-Cross}}]{Hashemi2014}%
  \BibitemOpen
  \bibfield  {author} {\bibinfo {author} {\bibfnamefont {R.}~\bibnamefont {Hashemi}}, \bibinfo {author} {\bibfnamefont {C.}~\bibnamefont {Povey}}, \bibinfo {author} {\bibfnamefont {M.}~\bibnamefont {Derksen}}, \bibinfo {author} {\bibfnamefont {H.}~\bibnamefont {Naseri}}, \bibinfo {author} {\bibfnamefont {J.}~\bibnamefont {Garber}}, \ and\ \bibinfo {author} {\bibfnamefont {A.}~\bibnamefont {Predoi-Cross}},\ }\href {\doibase 10.1063/1.4902076} {\bibfield  {journal} {\bibinfo  {journal} {J. Chem. Phys.}\ }\textbf {\bibinfo {volume} {141}},\ \bibinfo {pages} {214201} (\bibinfo {year} {2014})}\BibitemShut {NoStop}%
\bibitem [{\citenamefont {Gotti}\ \emph {et~al.}(2018)\citenamefont {Gotti}, \citenamefont {Moretti}, \citenamefont {Gatti}, \citenamefont {Castrillo}, \citenamefont {Galzerano}, \citenamefont {Laporta}, \citenamefont {Gianfrani},\ and\ \citenamefont {Marangoni}}]{Gotti2018}%
  \BibitemOpen
  \bibfield  {author} {\bibinfo {author} {\bibfnamefont {R.}~\bibnamefont {Gotti}}, \bibinfo {author} {\bibfnamefont {L.}~\bibnamefont {Moretti}}, \bibinfo {author} {\bibfnamefont {D.}~\bibnamefont {Gatti}}, \bibinfo {author} {\bibfnamefont {A.}~\bibnamefont {Castrillo}}, \bibinfo {author} {\bibfnamefont {G.}~\bibnamefont {Galzerano}}, \bibinfo {author} {\bibfnamefont {P.}~\bibnamefont {Laporta}}, \bibinfo {author} {\bibfnamefont {L.}~\bibnamefont {Gianfrani}}, \ and\ \bibinfo {author} {\bibfnamefont {M.}~\bibnamefont {Marangoni}},\ }\href {\doibase 10.1103/PhysRevA.97.012512} {\bibfield  {journal} {\bibinfo  {journal} {Phys. Rev. A}\ }\textbf {\bibinfo {volume} {97}},\ \bibinfo {pages} {012512} (\bibinfo {year} {2018})}\BibitemShut {NoStop}%
\bibitem [{\citenamefont {Truong}\ \emph {et~al.}(2015)\citenamefont {Truong}, \citenamefont {Anstie}, \citenamefont {May}, \citenamefont {Stace},\ and\ \citenamefont {Luiten}}]{Truong2015}%
  \BibitemOpen
  \bibfield  {author} {\bibinfo {author} {\bibfnamefont {G.-W.}\ \bibnamefont {Truong}}, \bibinfo {author} {\bibfnamefont {J.~D.}\ \bibnamefont {Anstie}}, \bibinfo {author} {\bibfnamefont {E.~F.}\ \bibnamefont {May}}, \bibinfo {author} {\bibfnamefont {T.~M.}\ \bibnamefont {Stace}}, \ and\ \bibinfo {author} {\bibfnamefont {A.~N.}\ \bibnamefont {Luiten}},\ }\href {\doibase 10.1038/ncomms9345} {\bibfield  {journal} {\bibinfo  {journal} {Nat. Commun.}\ }\textbf {\bibinfo {volume} {6}},\ \bibinfo {pages} {8343} (\bibinfo {year} {2015})}\BibitemShut {NoStop}%
\bibitem [{\citenamefont {Gatti}\ \emph {et~al.}(2013)\citenamefont {Gatti}, \citenamefont {Mills}, \citenamefont {De~Vizia}, \citenamefont {Mohr}, \citenamefont {Hartl}, \citenamefont {Marangoni}, \citenamefont {Fermann},\ and\ \citenamefont {Gianfrani}}]{Gatti2013}%
  \BibitemOpen
  \bibfield  {author} {\bibinfo {author} {\bibfnamefont {D.}~\bibnamefont {Gatti}}, \bibinfo {author} {\bibfnamefont {A.~A.}\ \bibnamefont {Mills}}, \bibinfo {author} {\bibfnamefont {M.~D.}\ \bibnamefont {De~Vizia}}, \bibinfo {author} {\bibfnamefont {C.}~\bibnamefont {Mohr}}, \bibinfo {author} {\bibfnamefont {I.}~\bibnamefont {Hartl}}, \bibinfo {author} {\bibfnamefont {M.}~\bibnamefont {Marangoni}}, \bibinfo {author} {\bibfnamefont {M.}~\bibnamefont {Fermann}}, \ and\ \bibinfo {author} {\bibfnamefont {L.}~\bibnamefont {Gianfrani}},\ }\href {\doibase 10.1103/PhysRevA.88.012514} {\bibfield  {journal} {\bibinfo  {journal} {Phys. Rev. A}\ }\textbf {\bibinfo {volume} {88}},\ \bibinfo {pages} {012514} (\bibinfo {year} {2013})}\BibitemShut {NoStop}%
\bibitem [{\citenamefont {Castrillo}\ \emph {et~al.}(2019)\citenamefont {Castrillo}, \citenamefont {Fasci}, \citenamefont {Dinesan}, \citenamefont {Gravina}, \citenamefont {Moretti},\ and\ \citenamefont {Gianfrani}}]{Castrillo2019}%
  \BibitemOpen
  \bibfield  {author} {\bibinfo {author} {\bibfnamefont {A.}~\bibnamefont {Castrillo}}, \bibinfo {author} {\bibfnamefont {E.}~\bibnamefont {Fasci}}, \bibinfo {author} {\bibfnamefont {H.}~\bibnamefont {Dinesan}}, \bibinfo {author} {\bibfnamefont {S.}~\bibnamefont {Gravina}}, \bibinfo {author} {\bibfnamefont {L.}~\bibnamefont {Moretti}}, \ and\ \bibinfo {author} {\bibfnamefont {L.}~\bibnamefont {Gianfrani}},\ }\href {\doibase 10.1103/PhysRevApplied.11.064060} {\bibfield  {journal} {\bibinfo  {journal} {Phys. Rev. Appl.}\ }\textbf {\bibinfo {volume} {11}},\ \bibinfo {pages} {064060} (\bibinfo {year} {2019})}\BibitemShut {NoStop}%
\bibitem [{\citenamefont {Hodges}\ \emph {et~al.}(2025)\citenamefont {Hodges}, \citenamefont {Bielska}, \citenamefont {Birk}, \citenamefont {Guo}, \citenamefont {Li}, \citenamefont {Lim}, \citenamefont {Lisak}, \citenamefont {Reed},\ and\ \citenamefont {Wagner}}]{CCQMP229}%
  \BibitemOpen
  \bibfield  {author} {\bibinfo {author} {\bibfnamefont {J.~T.}\ \bibnamefont {Hodges}}, \bibinfo {author} {\bibfnamefont {K.}~\bibnamefont {Bielska}}, \bibinfo {author} {\bibfnamefont {M.}~\bibnamefont {Birk}}, \bibinfo {author} {\bibfnamefont {R.}~\bibnamefont {Guo}}, \bibinfo {author} {\bibfnamefont {G.}~\bibnamefont {Li}}, \bibinfo {author} {\bibfnamefont {J.~S.}\ \bibnamefont {Lim}}, \bibinfo {author} {\bibfnamefont {D.}~\bibnamefont {Lisak}}, \bibinfo {author} {\bibfnamefont {Z.~D.}\ \bibnamefont {Reed}}, \ and\ \bibinfo {author} {\bibfnamefont {G.}~\bibnamefont {Wagner}},\ }\href {\doibase 10.1088/0026-1394/62/1A/08006} {\bibfield  {journal} {\bibinfo  {journal} {Metrologia}\ }\textbf {\bibinfo {volume} {62}},\ \bibinfo {pages} {08006} (\bibinfo {year} {2025})}\BibitemShut {NoStop}%
\bibitem [{\citenamefont {Huang}\ \emph {et~al.}(2024)\citenamefont {Huang}, \citenamefont {Tan}, \citenamefont {Yin}, \citenamefont {Nie}, \citenamefont {Wang},\ and\ \citenamefont {Hu}}]{Huang2024}%
  \BibitemOpen
  \bibfield  {author} {\bibinfo {author} {\bibfnamefont {Q.}~\bibnamefont {Huang}}, \bibinfo {author} {\bibfnamefont {Y.}~\bibnamefont {Tan}}, \bibinfo {author} {\bibfnamefont {R.-H.}\ \bibnamefont {Yin}}, \bibinfo {author} {\bibfnamefont {Z.-L.}\ \bibnamefont {Nie}}, \bibinfo {author} {\bibfnamefont {J.}~\bibnamefont {Wang}}, \ and\ \bibinfo {author} {\bibfnamefont {S.-M.}\ \bibnamefont {Hu}},\ }\href {\doibase 10.1088/1681-7575/ad7ec0} {\bibfield  {journal} {\bibinfo  {journal} {Metrologia}\ }\textbf {\bibinfo {volume} {61}},\ \bibinfo {pages} {065003} (\bibinfo {year} {2024})}\BibitemShut {NoStop}%
\bibitem [{\citenamefont {Cygan}\ \emph {et~al.}(2019)\citenamefont {Cygan}, \citenamefont {Wcis{\l}o}, \citenamefont {W\'{o}jtewicz}, \citenamefont {Kowzan}, \citenamefont {Zaborowski}, \citenamefont {Charczun}, \citenamefont {Bielska}, \citenamefont {Trawi\'{n}ski}, \citenamefont {Ciury{\l}o}, \citenamefont {Mas{\l}owski},\ and\ \citenamefont {Lisak}}]{Cygan2019}%
  \BibitemOpen
  \bibfield  {author} {\bibinfo {author} {\bibfnamefont {A.}~\bibnamefont {Cygan}}, \bibinfo {author} {\bibfnamefont {P.}~\bibnamefont {Wcis{\l}o}}, \bibinfo {author} {\bibfnamefont {S.}~\bibnamefont {W\'{o}jtewicz}}, \bibinfo {author} {\bibfnamefont {G.}~\bibnamefont {Kowzan}}, \bibinfo {author} {\bibfnamefont {M.}~\bibnamefont {Zaborowski}}, \bibinfo {author} {\bibfnamefont {D.}~\bibnamefont {Charczun}}, \bibinfo {author} {\bibfnamefont {K.}~\bibnamefont {Bielska}}, \bibinfo {author} {\bibfnamefont {R.~S.}\ \bibnamefont {Trawi\'{n}ski}}, \bibinfo {author} {\bibfnamefont {R.}~\bibnamefont {Ciury{\l}o}}, \bibinfo {author} {\bibfnamefont {P.}~\bibnamefont {Mas{\l}owski}}, \ and\ \bibinfo {author} {\bibfnamefont {D.}~\bibnamefont {Lisak}},\ }\href {\doibase 10.1364/OE.27.021810} {\bibfield  {journal} {\bibinfo  {journal} {Opt. Express}\ }\textbf {\bibinfo {volume} {27}},\ \bibinfo {pages} {21810} (\bibinfo {year} {2019})}\BibitemShut {NoStop}%
\bibitem [{\citenamefont {Casa}\ \emph {et~al.}(2008)\citenamefont {Casa}, \citenamefont {Castrillo}, \citenamefont {Galzerano}, \citenamefont {Wehr}, \citenamefont {Merlone}, \citenamefont {Di~Serafino}, \citenamefont {Laporta},\ and\ \citenamefont {Gianfrani}}]{Casa2008}%
  \BibitemOpen
  \bibfield  {author} {\bibinfo {author} {\bibfnamefont {G.}~\bibnamefont {Casa}}, \bibinfo {author} {\bibfnamefont {A.}~\bibnamefont {Castrillo}}, \bibinfo {author} {\bibfnamefont {G.}~\bibnamefont {Galzerano}}, \bibinfo {author} {\bibfnamefont {R.}~\bibnamefont {Wehr}}, \bibinfo {author} {\bibfnamefont {A.}~\bibnamefont {Merlone}}, \bibinfo {author} {\bibfnamefont {D.}~\bibnamefont {Di~Serafino}}, \bibinfo {author} {\bibfnamefont {P.}~\bibnamefont {Laporta}}, \ and\ \bibinfo {author} {\bibfnamefont {L.}~\bibnamefont {Gianfrani}},\ }\href {\doibase 10.1103/PhysRevLett.100.200801} {\bibfield  {journal} {\bibinfo  {journal} {Phys. Rev. Lett.}\ }\textbf {\bibinfo {volume} {100}},\ \bibinfo {pages} {200801} (\bibinfo {year} {2008})}\BibitemShut {NoStop}%
\bibitem [{\citenamefont {Yamada}\ \emph {et~al.}(2008)\citenamefont {Yamada}, \citenamefont {Onae}, \citenamefont {Hong}, \citenamefont {Inaba}, \citenamefont {Matsumoto}, \citenamefont {Nakajima}, \citenamefont {Ito},\ and\ \citenamefont {Shimizu}}]{Yamada2008}%
  \BibitemOpen
  \bibfield  {author} {\bibinfo {author} {\bibfnamefont {K.}~\bibnamefont {Yamada}}, \bibinfo {author} {\bibfnamefont {A.}~\bibnamefont {Onae}}, \bibinfo {author} {\bibfnamefont {F.-L.}\ \bibnamefont {Hong}}, \bibinfo {author} {\bibfnamefont {H.}~\bibnamefont {Inaba}}, \bibinfo {author} {\bibfnamefont {H.}~\bibnamefont {Matsumoto}}, \bibinfo {author} {\bibfnamefont {Y.}~\bibnamefont {Nakajima}}, \bibinfo {author} {\bibfnamefont {F.}~\bibnamefont {Ito}}, \ and\ \bibinfo {author} {\bibfnamefont {T.}~\bibnamefont {Shimizu}},\ }\href {\doibase https://doi.org/10.1016/j.jms.2008.03.002} {\bibfield  {journal} {\bibinfo  {journal} {Journal of Molecular Spectroscopy}\ }\textbf {\bibinfo {volume} {249}},\ \bibinfo {pages} {95} (\bibinfo {year} {2008})}\BibitemShut {NoStop}%
\bibitem [{\citenamefont {Lemarchand}\ \emph {et~al.}(2011)\citenamefont {Lemarchand}, \citenamefont {Triki}, \citenamefont {Darquié}, \citenamefont {Bordé}, \citenamefont {Chardonnet},\ and\ \citenamefont {Daussy}}]{Lemarchand2011}%
  \BibitemOpen
  \bibfield  {author} {\bibinfo {author} {\bibfnamefont {C.}~\bibnamefont {Lemarchand}}, \bibinfo {author} {\bibfnamefont {M.}~\bibnamefont {Triki}}, \bibinfo {author} {\bibfnamefont {B.}~\bibnamefont {Darquié}}, \bibinfo {author} {\bibfnamefont {C.~J.}\ \bibnamefont {Bordé}}, \bibinfo {author} {\bibfnamefont {C.}~\bibnamefont {Chardonnet}}, \ and\ \bibinfo {author} {\bibfnamefont {C.}~\bibnamefont {Daussy}},\ }\href {\doibase 10.1088/1367-2630/13/7/073028} {\bibfield  {journal} {\bibinfo  {journal} {New Journal of Physics}\ }\textbf {\bibinfo {volume} {13}},\ \bibinfo {pages} {073028} (\bibinfo {year} {2011})}\BibitemShut {NoStop}%
\bibitem [{\citenamefont {Truong}\ \emph {et~al.}(2011)\citenamefont {Truong}, \citenamefont {May}, \citenamefont {Stace},\ and\ \citenamefont {Luiten}}]{Truong2011}%
  \BibitemOpen
  \bibfield  {author} {\bibinfo {author} {\bibfnamefont {G.-W.}\ \bibnamefont {Truong}}, \bibinfo {author} {\bibfnamefont {E.~F.}\ \bibnamefont {May}}, \bibinfo {author} {\bibfnamefont {T.~M.}\ \bibnamefont {Stace}}, \ and\ \bibinfo {author} {\bibfnamefont {A.~N.}\ \bibnamefont {Luiten}},\ }\href {\doibase 10.1103/PhysRevA.83.033805} {\bibfield  {journal} {\bibinfo  {journal} {Phys. Rev. A}\ }\textbf {\bibinfo {volume} {83}},\ \bibinfo {pages} {033805} (\bibinfo {year} {2011})}\BibitemShut {NoStop}%
\bibitem [{\citenamefont {Cheng}\ \emph {et~al.}(2015)\citenamefont {Cheng}, \citenamefont {Wang}, \citenamefont {Sun}, \citenamefont {Tan}, \citenamefont {Kang},\ and\ \citenamefont {Hu}}]{Cheng2015}%
  \BibitemOpen
  \bibfield  {author} {\bibinfo {author} {\bibfnamefont {C.-F.}\ \bibnamefont {Cheng}}, \bibinfo {author} {\bibfnamefont {J.}~\bibnamefont {Wang}}, \bibinfo {author} {\bibfnamefont {Y.~R.}\ \bibnamefont {Sun}}, \bibinfo {author} {\bibfnamefont {Y.}~\bibnamefont {Tan}}, \bibinfo {author} {\bibfnamefont {P.}~\bibnamefont {Kang}}, \ and\ \bibinfo {author} {\bibfnamefont {S.-M.}\ \bibnamefont {Hu}},\ }\href {\doibase 10.1088/0026-1394/52/5/S385} {\bibfield  {journal} {\bibinfo  {journal} {Metrologia}\ }\textbf {\bibinfo {volume} {52}},\ \bibinfo {pages} {S385} (\bibinfo {year} {2015})}\BibitemShut {NoStop}%
\bibitem [{\citenamefont {Galzerano}(2020)}]{Galzerano2020}%
  \BibitemOpen
  \bibfield  {author} {\bibinfo {author} {\bibfnamefont {G.}~\bibnamefont {Galzerano}},\ }\href {\doibase https://doi.org/10.1016/j.measurement.2020.107940} {\bibfield  {journal} {\bibinfo  {journal} {Measurement}\ }\textbf {\bibinfo {volume} {164}},\ \bibinfo {pages} {107940} (\bibinfo {year} {2020})}\BibitemShut {NoStop}%
\bibitem [{\citenamefont {Wehr}\ \emph {et~al.}(2003)\citenamefont {Wehr}, \citenamefont {McKernan}, \citenamefont {Vitcu}, \citenamefont {Ciurylo},\ and\ \citenamefont {Drummond}}]{Wehr2003}%
  \BibitemOpen
  \bibfield  {author} {\bibinfo {author} {\bibfnamefont {R.}~\bibnamefont {Wehr}}, \bibinfo {author} {\bibfnamefont {E.}~\bibnamefont {McKernan}}, \bibinfo {author} {\bibfnamefont {A.}~\bibnamefont {Vitcu}}, \bibinfo {author} {\bibfnamefont {R.}~\bibnamefont {Ciurylo}}, \ and\ \bibinfo {author} {\bibfnamefont {J.~R.}\ \bibnamefont {Drummond}},\ }\href {\doibase 10.1364/AO.42.006595} {\bibfield  {journal} {\bibinfo  {journal} {Appl. Opt.}\ }\textbf {\bibinfo {volume} {42}},\ \bibinfo {pages} {6595} (\bibinfo {year} {2003})}\BibitemShut {NoStop}%
\bibitem [{\citenamefont {Tennyson}\ \emph {et~al.}(2014)\citenamefont {Tennyson}, \citenamefont {Bernath}, \citenamefont {Campargue}, \citenamefont {Császár}, \citenamefont {Daumont}, \citenamefont {Gamache}, \citenamefont {Hodges}, \citenamefont {Lisak}, \citenamefont {Naumenko}, \citenamefont {Rothman}, \citenamefont {Tran}, \citenamefont {Zobov}, \citenamefont {Buldyreva}, \citenamefont {Boone}, \citenamefont {Vizia}, \citenamefont {Gianfrani}, \citenamefont {Hartmann}, \citenamefont {McPheat}, \citenamefont {Weidmann}, \citenamefont {Murray}, \citenamefont {Ngo},\ and\ \citenamefont {Polyansky}}]{Tennyson2014}%
  \BibitemOpen
  \bibfield  {author} {\bibinfo {author} {\bibfnamefont {J.}~\bibnamefont {Tennyson}}, \bibinfo {author} {\bibfnamefont {P.~F.}\ \bibnamefont {Bernath}}, \bibinfo {author} {\bibfnamefont {A.}~\bibnamefont {Campargue}}, \bibinfo {author} {\bibfnamefont {A.~G.}\ \bibnamefont {Császár}}, \bibinfo {author} {\bibfnamefont {L.}~\bibnamefont {Daumont}}, \bibinfo {author} {\bibfnamefont {R.~R.}\ \bibnamefont {Gamache}}, \bibinfo {author} {\bibfnamefont {J.~T.}\ \bibnamefont {Hodges}}, \bibinfo {author} {\bibfnamefont {D.}~\bibnamefont {Lisak}}, \bibinfo {author} {\bibfnamefont {O.~V.}\ \bibnamefont {Naumenko}}, \bibinfo {author} {\bibfnamefont {L.~S.}\ \bibnamefont {Rothman}}, \bibinfo {author} {\bibfnamefont {H.}~\bibnamefont {Tran}}, \bibinfo {author} {\bibfnamefont {N.~F.}\ \bibnamefont {Zobov}}, \bibinfo {author} {\bibfnamefont {J.}~\bibnamefont {Buldyreva}}, \bibinfo {author} {\bibfnamefont {C.~D.}\ \bibnamefont {Boone}}, \bibinfo {author} {\bibfnamefont {M.~D.~D.}\ \bibnamefont {Vizia}}, \bibinfo {author}
  {\bibfnamefont {L.}~\bibnamefont {Gianfrani}}, \bibinfo {author} {\bibfnamefont {J.-M.}\ \bibnamefont {Hartmann}}, \bibinfo {author} {\bibfnamefont {R.}~\bibnamefont {McPheat}}, \bibinfo {author} {\bibfnamefont {D.}~\bibnamefont {Weidmann}}, \bibinfo {author} {\bibfnamefont {J.}~\bibnamefont {Murray}}, \bibinfo {author} {\bibfnamefont {N.~H.}\ \bibnamefont {Ngo}}, \ and\ \bibinfo {author} {\bibfnamefont {O.~L.}\ \bibnamefont {Polyansky}},\ }\href {\doibase doi:10.1515/pac-2014-0208} {\bibfield  {journal} {\bibinfo  {journal} {Pure Appl. Chem.}\ }\textbf {\bibinfo {volume} {86}},\ \bibinfo {pages} {1931} (\bibinfo {year} {2014})}\BibitemShut {NoStop}%
\bibitem [{\citenamefont {Bousheheri}\ \emph {et~al.}(1987)\citenamefont {Bousheheri}, \citenamefont {Bzowski}, \citenamefont {Kestin},\ and\ \citenamefont {Mason}}]{Bousheheri1987}%
  \BibitemOpen
  \bibfield  {author} {\bibinfo {author} {\bibfnamefont {A.}~\bibnamefont {Bousheheri}}, \bibinfo {author} {\bibfnamefont {J.}~\bibnamefont {Bzowski}}, \bibinfo {author} {\bibfnamefont {J.}~\bibnamefont {Kestin}}, \ and\ \bibinfo {author} {\bibfnamefont {E.~A.}\ \bibnamefont {Mason}},\ }\href {\doibase 10.1063/1.555800} {\bibfield  {journal} {\bibinfo  {journal} {J. Phys. Chem. Ref. Data}\ }\textbf {\bibinfo {volume} {16}},\ \bibinfo {pages} {445} (\bibinfo {year} {1987})}\BibitemShut {NoStop}%
\bibitem [{\citenamefont {Drever}\ \emph {et~al.}(1983)\citenamefont {Drever}, \citenamefont {Hall}, \citenamefont {Kowalski}, \citenamefont {Hough}, \citenamefont {Ford}, \citenamefont {Munley},\ and\ \citenamefont {Ward}}]{Drever1983}%
  \BibitemOpen
  \bibfield  {author} {\bibinfo {author} {\bibfnamefont {R.~W.~P.}\ \bibnamefont {Drever}}, \bibinfo {author} {\bibfnamefont {J.~L.}\ \bibnamefont {Hall}}, \bibinfo {author} {\bibfnamefont {F.~V.}\ \bibnamefont {Kowalski}}, \bibinfo {author} {\bibfnamefont {J.}~\bibnamefont {Hough}}, \bibinfo {author} {\bibfnamefont {G.~M.}\ \bibnamefont {Ford}}, \bibinfo {author} {\bibfnamefont {A.~J.}\ \bibnamefont {Munley}}, \ and\ \bibinfo {author} {\bibfnamefont {H.}~\bibnamefont {Ward}},\ }\href {\doibase 10.1007/BF00702605} {\bibfield  {journal} {\bibinfo  {journal} {Appl. Phys. B}\ }\textbf {\bibinfo {volume} {31}},\ \bibinfo {pages} {97} (\bibinfo {year} {1983})}\BibitemShut {NoStop}%
\bibitem [{\citenamefont {Cygan}\ \emph {et~al.}(2016)\citenamefont {Cygan}, \citenamefont {W\'{o}jtewicz}, \citenamefont {Kowzan}, \citenamefont {Zaborowski}, \citenamefont {Wcis{\l}o}, \citenamefont {Nawrocki}, \citenamefont {Krehlik}, \citenamefont {\'{S}liwczyński}, \citenamefont {Lipi\'{n}ski}, \citenamefont {Mas{\l}owski}, \citenamefont {Ciury{\l}o},\ and\ \citenamefont {Lisak}}]{Cygan2016}%
  \BibitemOpen
  \bibfield  {author} {\bibinfo {author} {\bibfnamefont {A.}~\bibnamefont {Cygan}}, \bibinfo {author} {\bibfnamefont {S.}~\bibnamefont {W\'{o}jtewicz}}, \bibinfo {author} {\bibfnamefont {G.}~\bibnamefont {Kowzan}}, \bibinfo {author} {\bibfnamefont {M.}~\bibnamefont {Zaborowski}}, \bibinfo {author} {\bibfnamefont {P.}~\bibnamefont {Wcis{\l}o}}, \bibinfo {author} {\bibfnamefont {J.}~\bibnamefont {Nawrocki}}, \bibinfo {author} {\bibfnamefont {P.}~\bibnamefont {Krehlik}}, \bibinfo {author} {\bibfnamefont {{\L}.}~\bibnamefont {\'{S}liwczyński}}, \bibinfo {author} {\bibfnamefont {M.}~\bibnamefont {Lipi\'{n}ski}}, \bibinfo {author} {\bibfnamefont {P.}~\bibnamefont {Mas{\l}owski}}, \bibinfo {author} {\bibfnamefont {R.}~\bibnamefont {Ciury{\l}o}}, \ and\ \bibinfo {author} {\bibfnamefont {D.}~\bibnamefont {Lisak}},\ }\href {\doibase 10.1063/1.4952651} {\bibfield  {journal} {\bibinfo  {journal} {J. Chem. Phys.}\ }\textbf {\bibinfo {volume} {144}},\ \bibinfo {pages} {214202} (\bibinfo {year} {2016})}\BibitemShut
  {NoStop}%
\end{thebibliography}


\providecommand{\noopsort}[1]{}\providecommand{\singleletter}[1]{#1}%
%



\appendix
\section{Expression for temperature }
\label{appxi}

In Eq. (\ref{EqT2L}), the quantity $\xi(T)$ depends on temperature. In general:
\begin{equation}
    \label{EqxiT}
    \xi(T)= R_S(T_r) 
    e^{-\frac{\Delta E''}{k T_r}}
    \frac{1-e^{-\frac{h\nu_a}{k T_r}}}
    {1-e^{-\frac{h\nu_b}{k T_r}}}
    \frac{1-e^{-\frac{h\nu_b}{k T}}}
    {1-e^{-\frac{h\nu_a}{k T}}}
    \; ,
\end{equation}
where $\nu_a$ and $\nu_b$ are frequencies of used lines, and $h$ is the Planck constant.
However, in the case when $h\nu_a/(k T)$ and $h\nu_b/(k T)$ are much less than unity, this temperature dependence can be neglected.

\section{Non-impact and line-mixing effects}
\label{appNILM}

The non-impact effects \cite{Boulet2004, Reed2023}, as well as line-mixing going beyond first-order approximation \cite{Ciurylo2001LM, Pine2019}, can lead to a nonlinear variation of apparent line intensity on gas pressure. However, in the case of the study presented in this work, the estimated relative error of measured intensity caused by neglecting these effects is below $10^{-4}$. Moreover, this conclusion is supported by results discussed in Appendix \ref{app2}, where no departures from line area linearity with pressure were observed within experimental uncertainty when the pressure range was varied.

\section{Experimental setup}
\label{app1}

\begin{figure*}[t]
\centering
\includegraphics[width=\textwidth]{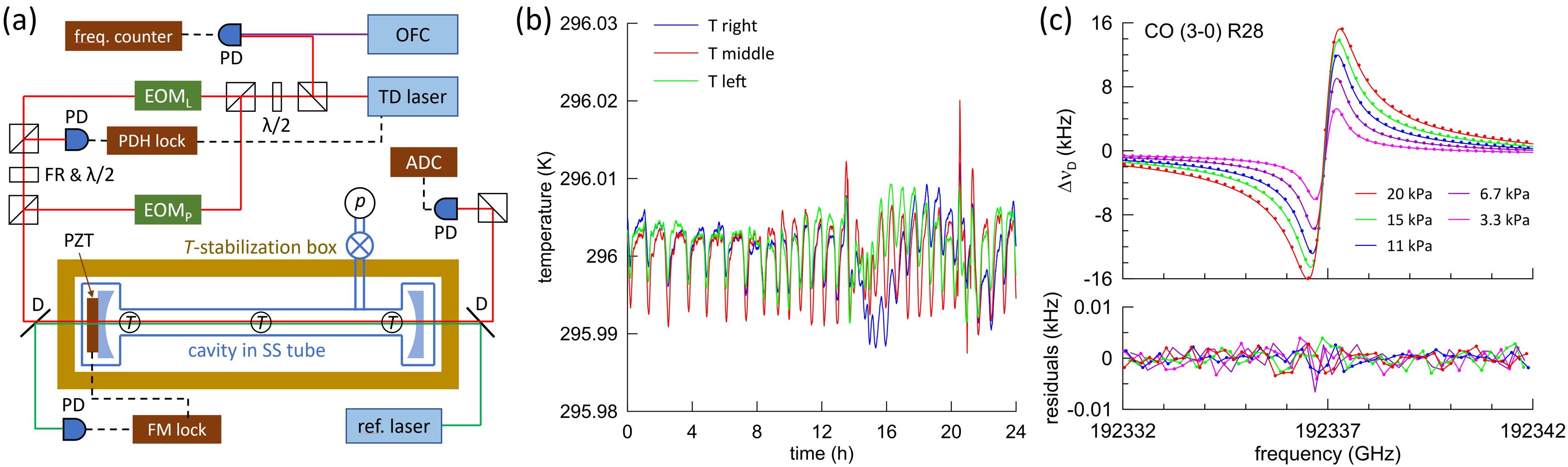}
\caption{(a) Experimental setup of CMDS spectrometer: EOM - electro-optic modulator, ADC - analog-to-digital converter, TD laser - tunable diode laser, OFC - optical frequency comb, ref. laser - reference Nd:YAG laser, PD - photodetector, D - dichroic mirror, FR - Faraday rotator, $\lambda$/2 - half-wave plate, $T$ - temperature sensors, $p$ - pressure sensor, FM lock - frequency-modulation locking circuit, PZT - piezo-electric mirror actuator, PDH lock - Pound-Drever-Hall locking circuit. (b) Temperature stability of the cavity in time, measured at three points along the tube containing the gas sample. (c) Example spectra of CO (3-0) band line R28 at five pressures and the residuals from the fitted HTP.} \label{Fig_ExperimSetup}
\end{figure*}

The CMDS spectrometer is schematically shown in Fig. \ref{Fig_ExperimSetup}(a). 
The tunable external cavity diode laser is frequency-locked to the high-finesse (${\mathcal F}=41000$) cavity with the sample CO gas inside, using the Pound-Drever-Hall \cite{Drever1983} technique. The orthogonally polarized probe beam, separated from the laser output, is scanned through consecutive cavity modes using a 20-GHz-bandwidth electro-optic modulator to measure the mode center frequency differences from the mode to which the laser is locked. 
The cavity has a free spectral range of about 202 MHz and has an optical path length locked to the I$_2$-stabilized Nd:YAG laser, leading to the cavity mode absolute frequency stability, measured using optical frequency comb, below 10 kHz and a relative laser-to-cavity line width below 40 Hz \cite{Cygan2016}. 

The cavity is placed in a thermal insulation box with active temperature control. Temperatures, measured with three 10-k$\Omega$ thermistors, are shown in Fig. \ref{Fig_ExperimSetup}(b). The thermistors were calibrated to the reference-grade thermometer ({\it Fluke 1595A + 5641}) calibrated to SI with an uncertainty of 2 mK near 296 K. The calibration process, temperature stability, and uniformity led to combined standard uncertainty of the cavity temperature, $u(T)=30$ mK. The gas pressure was measured with capacitance manometers calibrated to SI with a relative combined standard uncertainty $u(p)/p = 0.1$\% to 0.008\% for pressures of 1.3 to 20 kPa ({\it Wika CPG2500 + CPR2580}) and $u(p)/p= 0.05$\% of reading for pressures 20 Pa to 1.3~kPa ({\it MKS Baratron 690A}).

An example spectra of line R28 from the CO (3-0) band, measured at a temperature of  296.00(3)~K and CO pressures between 3.3 kPa and 20 kPa, are shown in Fig. \ref{Fig_ExperimSetup}(c). The fit residuals from the Hartmann-Tran profile (HTP) \cite{Ngo2013, Tennyson2014} show agreement of this model with the spectra having signal-to-noise ratio between 6000 and 19000. 

\begin{figure}
\centering
\includegraphics[width=0.48\textwidth]{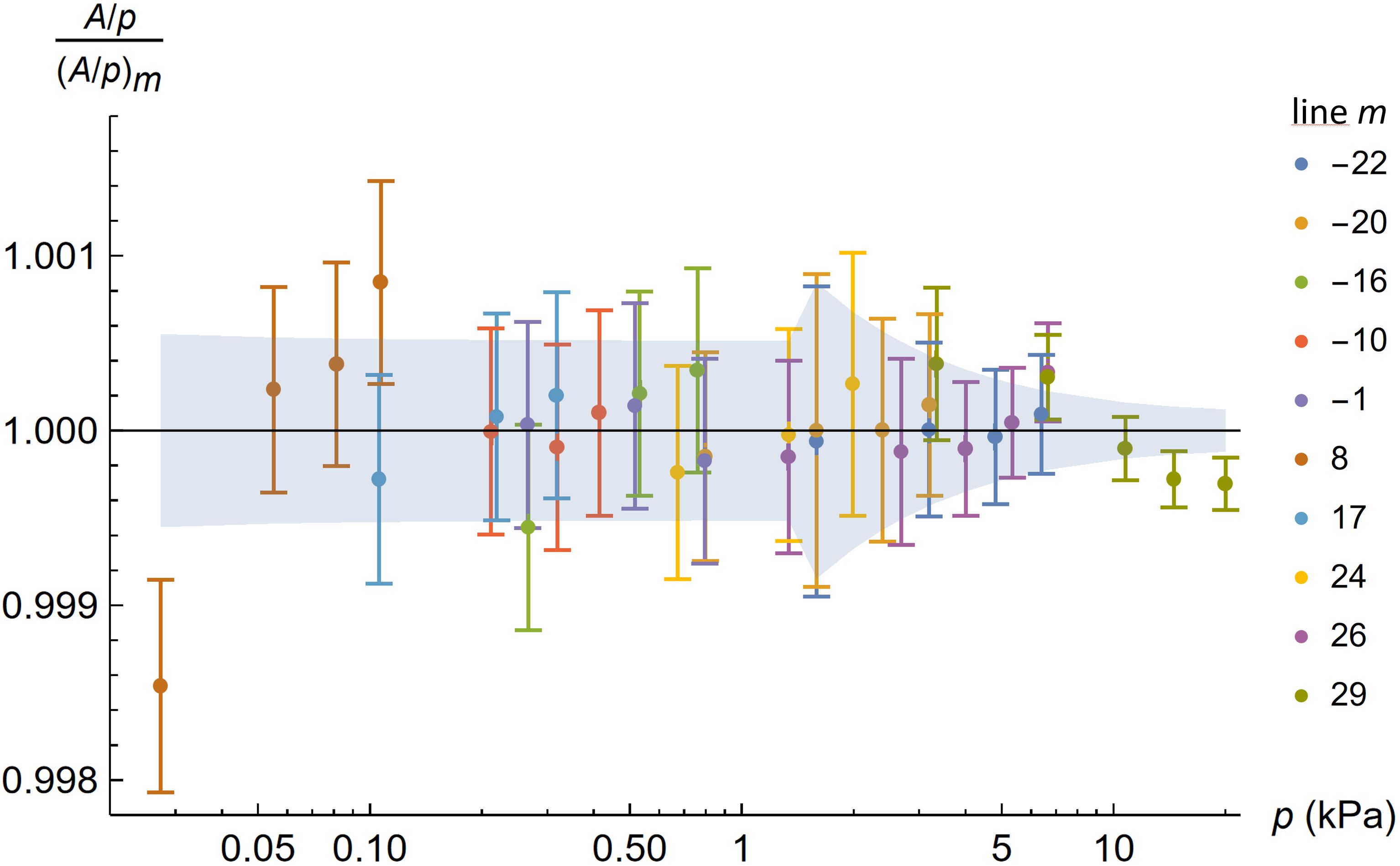}
\caption{The ratios of the fitted line area divided by the pressure $(A/p)$ to its mean value $(A/p)_m$ for all pressures at which the given line was measured plotted as a function of gas pressure. Point colors indicate the CO line index $m$. Error bars correspond to combined standard uncertainties. The gray area corresponds to the standard uncertainty of gas concentration measurements using calibrated $p$ and $T$ sensors.} \label{Fig_pLin}
\end{figure}

\section{Linearity of pressure gauge}
\label{app2}

We verified the linearity of our pressure gauge spectroscopically by comparison of the integrated line areas, $A$, and the pressure gauge readings, $p$. In Fig. \ref{Fig_pLin}, we show the ratios $(A/p)/(A/p)_m$, where $(A/p)_m$ is the mean value for all pressures at which the given line was measured. The mean value of standard deviations of these ratios for all lines is 0.030\%. The low-end point at 27 Pa, deviating from others by about 0.2\%, shows the limitation of the high-linearity range down to about 50 Pa. For $p$ between 50 Pa and 20 kPa, the standard deviation of the $(A/p)/(A/p)_m$ is 0.023\%.

\end{document}